\documentclass{article}

\usepackage{graphicx}
\usepackage{xcolor}
\usepackage{amsmath, xparse}
\usepackage{amsfonts}
\usepackage{amssymb}
\usepackage{breqn}
\usepackage{acro}
\usepackage{setspace}
\usepackage{authblk}
\usepackage{hyperref}
\usepackage{subcaption}
\usepackage{braket}
\usepackage{siunitx}
\usepackage{physics}
\usepackage{marginnote}
\usepackage{multirow}
\usepackage{float}
\usepackage{longtable}
\usepackage[color=orange!60,textsize=scriptsize]{todonotes}
\usepackage[style=chem-acs,]{biblatex}
\usepackage{booktabs}
\usepackage{xr-hyper}
\usepackage{hyperref}

\AtBeginDocument{%
\heavyrulewidth=.08em
\lightrulewidth=.05em
\cmidrulewidth=.03em
\belowrulesep=.65ex
\belowbottomsep=0pt
\aboverulesep=.4ex
\abovetopsep=0pt
\cmidrulesep=\doublerulesep
\cmidrulekern=.5em
\defaultaddspace=.5em
}

\DeclareSIUnit\angstrom{\text{$\angstrom$}} 
\DeclareSourcemap{
  \maps[datatype=bibtex]{
    \map{
      \step[fieldsource=month, match=\regexp{(?i)jan(uary)?}, replace={01}]
      \step[fieldsource=month, match=\regexp{(?i)feb(ruary)?}, replace={02}]
      \step[fieldsource=month, match=\regexp{(?i)mar(ch)?}, replace={03}]
      \step[fieldsource=month, match=\regexp{(?i)apr(il)?}, replace={04}]
      \step[fieldsource=month, match=\regexp{(?i)may}, replace={05}]
      \step[fieldsource=month, match=\regexp{(?i)jun(e)?}, replace={06}]
      \step[fieldsource=month, match=\regexp{(?i)jul(y)?}, replace={07}]
      \step[fieldsource=month, match=\regexp{(?i)aug(ust)?}, replace={08}]
      \step[fieldsource=month, match=\regexp{(?i)sep(t(ember)?)?}, replace={09}]
      \step[fieldsource=month, match=\regexp{(?i)oct(ober)?}, replace={10}]
      \step[fieldsource=month, match=\regexp{(?i)nov(ember)?}, replace={11}]
      \step[fieldsource=month, match=\regexp{(?i)dec(ember)?}, replace={12}]
    }
  }
}

\newcommand{\angstrom}{\textup{\AA}}

\graphicspath{{../images/}}

\author[1]{Shahzad Akram}
\author[2]{Sutirtha Paul}
\author[3]{Collin Kovacs}
\author[3]{Vasileios Maroulas}
\author[2,4]{Adrian Del Maestro\thanks{Corresponding author: \href{mailto:Adrian.DelMaestro@utk.edu}{Adrian.DelMaestro@utk.edu}}}
\author[1]{Konstantinos D. Vogiatzis\thanks{Corresponding author: \href{mailto:kvogiatz@utk.edu}{kvogiatz@utk.edu}}}

\affil[1]{Department of Chemistry, University of Tennessee, Knoxville, TN, 37996, USA}
\affil[2]{Department of Physics and Astronomy, University of Tennessee, Knoxville, TN, 37996, USA}
\affil[3]{Department of Mathematics, University of Tennessee, Knoxville, TN, 37996, USA}
\affil[4]{Min. H. Kao Department of Electrical Engineering and Computer Science, University of Tennessee, Knoxville, TN, 37996, USA}

\date{\today}

\DeclareAcronym{cc}{
  short=CC,
  long=coupled-cluster,
}

\DeclareAcronym{ccsd}{
  short=CCSD,
  long=coupled-cluster singles and doubles
}

\DeclareAcronym{ccsd_t}{
  short=CCSD(T),
  long=coupled-cluster singles-and-doubles with perturbative triples
}

\DeclareAcronym{dft}{
   short=DFT,
   long=density functional theory
}

\DeclareAcronym{sapt}{
   short=SAPT,
   long = symmetry-adapted perturbation theory
}

\DeclareAcronym{fci}{
   short=FCI,
   long=full configuration interaction 
}

\DeclareAcronym{hf}{
   short=HF,
   long=Hartree-Fock
}

\DeclareAcronym{hebz}{
   short=He-Bz,
   long=helium-benzene
}

\DeclareAcronym{mlffs}{
   short=MLFFs,
   long=machine learning force fields
}

\DeclareAcronym{adz}{
  short=aug-cc-pVDZ,
  long=augmented correlation-consistent polarized valence double-zeta,
}

\DeclareAcronym{atz}{
  short=aug-cc-pVTZ,
  long=augmented correlation-consistent polarized valence triple-zeta,
}

\DeclareAcronym{aqz}{
  short=aug-cc-pVQZ,
  long=augmented correlation-consistent polarized valence quadruple-zeta,
}

\DeclareAcronym{cbs}{
  short=CBS,
  long=complete basis set,
}

\DeclareAcronym{cbs_dt}{
  short={aug-cc-pV(D,T)Z},
  long={aug-cc-pV(D,T)Z},
}

\DeclareAcronym{cbs_tq}{
  short={aug-cc-pV(T,Q)Z},
  long={aug-cc-pV(T,Q)Z},
}

\DeclareAcronym{rmse}{
  short=RMSE,
  long=mean squared error,
}

\DeclareAcronym{mae}{
  short=MAE,
  long=mean absolute error,
}

\DeclareAcronym{maxe}{
  short=MAXE,
  long=maximum error,
}

\DeclareAcronym{pahs}{
  short=PAHs,
  long= polycyclic aromatic hydrocarbons,
}

\DeclareAcronym{c_coronene}{
  short=c-coronene,
  long=circumcoronene,
}

\DeclareAcronym{cc_coronene}{
  short=cc-coronene,
  long=circumcircumcoronene,
}

\DeclareAcronym{ccc_coronene}{
  short=ccc-coronene,
  long=circumcircumcircumcoronene,
}

\DeclareAcronym{1_d}{
  short=1D,
  long=one-dimensional,
}

\DeclareAcronym{3_d}{
  short=3D,
  long=three-dimensional,
}

\DeclareAcronym{pimc}{
  short=PIMC,
  long=Path integral Monte Carlo,
}

\title{Accurate Helium-Benzene Potential: from CCSD(T) to Gaussian Process Regression}

\begin{document}

\maketitle

\pagebreak


\begin{spacing}{1.5}

\section*{Abstract}
The accurate modeling of non-covalent interactions between helium and graphitic materials is important for understanding quantum phenomena in reduced dimensions, with the helium-benzene complex serving as the fundamental prototype. However, creating a quantitatively reliable potential energy surface (PES) for this weakly bound system remains a significant computational challenge. In this work, we present a comprehensive, multi-level investigation of the He-Bz interaction, establishing benchmark energies using high-level coupled-cluster singles-and-doubles with perturbative triples (CCSD(T)) methods extrapolated to the complete basis set limit and assessing higher-order (CCSDT(Q)) contributions. We use symmetry-adapted perturbation theory (SAPT) to benchmark it against CCSD(T) and to decompose the interaction into its physical components—confirming it is dominated by a balance between dispersion and exchange-repulsion. A continuous, three-dimensional PES is constructed from discrete ab initio points using multifidelity Gaussian process regression that combines density functional theory results with sparse coupled-cluster energies. The result is a highly accurate PES with sub-\si{\centi\meter^{-1}} accuracy that obeys physical laws.  This new PES is applied to path integral Monte Carlo (PIMC) simulations to study the solvation of $^{4}\text{He}$ atoms on benzene at low temperatures. Our PIMC results reveal qualitatively different solvation behavior, particularly in the filling of adsorption layers, when compared to simulations using commonly employed empirical Lennard-Jones potentials. This work provides a benchmark PES essential for accurate many-body simulations of helium on larger polycyclic aromatic hydrocarbons towards graphene.

\pagebreak

\section{Introduction}
The accurate description of weak, non-covalent interactions is a central challenge in chemical physics. These long-range forces govern the structure, dynamics, and adsorption of molecules on surfaces, as well as molecular recognition in biochemical systems, substrate binding in catalysis, and self-assembly in supramolecular structures. \supercite{AlHamdani2021, serhii_tretiakov_nigam_pollice_2025, al-hamdani_tkatchenko_2019} The interaction of helium with carbon-based materials like graphene is a prime example. This system is of fundamental importance for understanding quantum phenomena in reduced dimensions, such as the formation of 2D superfluids and exotic adsorbed phases \supercite{Nichols2016, toennies2013helium, mudrich2014photoionisaton, yu_two-dimensional_2021,del_maestro_perspective_2022,kappe2022solvation, kim_strain-induced_2024}. Graphene, a monolayer of carbon atoms in a hexagonal lattice, has exceptional electronic, thermal, and mechanical properties and has a high surface area, \supercite{Urade_2022, Alves_2024, Guo_2021, Sahoo_2024} and its high surface area and conductivity can make it an excellent candidate for supporting helium films. The van der Waals interaction between helium and graphene can be tuned via mechanical strain, and an uniaxial strain and electron correlation effects can modulate the attractive forces between neutral adatoms and graphene. It suggests that mechanical strain can be used to engineer adsorption potentials, impacting the formation of anisotropic low-dimensional superfluid phases.\supercite{nichols_adsorption_2016,kim_strain-induced_2024} Graphene's polarizability enhances the van der Waals forces, promoting wetting and leading to critical thicknesses where liquid film growth is arrested, triggering surface instabilities and pattern formation.\supercite{sengupta_theory_2018} 


Despite substantial progress in theoretical and experimental investigations of the noncovalent interactions of helium \supercite{Kwon2001, heidenreich2001nonrigidity, Cappelletti2002, Huang2003, heidenreich2003permutational, felker2003intermolecular, Schmied2004, Vranje__2004, xu2007wave, gibbons2009quantum, whitley2009spectral, Rzepa_2010, whitley2011theoretical, Kievsky_2011, Bakr_2013, Cappelletti_2015, Borocci_2020, Gao_2020, Hayashi2020, Bacanu_2021, Zunzunegui_Bru_2022} a robust computational framework is essential to accurately describe the interactions between helium and larger \ac{pahs} like graphene. However, achieving high-accuracy simulations of helium adsorption on graphene remains computationally demanding. To make this problem computationally tractable, it makes more sense to start with some prototypical models. The \ac{hebz} complex serves as the essential building block—the minimal model for the helium-$\pi$ interaction. A quantitatively reliable potential energy surface (PES) for \ac{hebz} is the foundation for building accurate models for larger \ac{pahs} and, by extension, for graphene itself.

Computer simulations can provide a tunable framework for exploring solid-state phenomena, capturing exotic adsorbed phases such as two-dimensional superfluids and supersolids, and enabling more accurate modeling of helium–aromatic complex systems. For this purpose, accurate interaction potentials are required for the molecular complex under consideration. Despite its importance, a definitive \ac{hebz} potential has remained elusive. The interaction is exceptionally weak (on the order of a few cm$^{-1}$), making it notoriously difficult to model the full potential surface with fewer single-point calculations. Some theoretical studies have addressed the parameterization of the \ac{hebz} potential. Notably, Lee et al.\supercite{Lee_2003} utilized a dataset of 280 points calculated at the \ac{ccsd_t} level augmented with midbond basis functions \supercite{tao1992mo, cybulski1999ground}, which were fitted to a complex 18-parameter analytical functional form. Following a similar methodology, Shirkov et al.\supercite{Shirkov_2024} employed 349 data points from \ac{ccsd_t} and equation-of-motion coupled-cluster calculations for studying both ground and excited states of the \ac{hebz} system. However, a significant limitation of these approaches is the high complexity of their respective analytical functions, which hinders least-squares fitting, straightforward implementation, and reproducibility.

In this work, we develop a quantitatively reliable interaction model for the helium–benzene system by integrating high-level electronic structure theory calculations\cite{Townsend2019} with a physically informed machine-learning framework. Accurate CCSD(T) and CCSDT(Q) calculations extrapolated to the complete basis set limit establish a robust reference description of the interaction across configuration space, with symmetry-adapted perturbation theory (SAPT) providing a transparent physical interpretation that highlights the dominant role of dispersion balanced by short-range exchange repulsion (Section \ref{sec:cc}). Building on this reference data, we construct a continuous three-dimensional potential energy surface by introducing a multifidelity Gaussian process approach that systematically combines sparse, high-accuracy coupled-cluster data with dense, low-cost DFT calculations through a constrained kernel decomposition, allowing shared spatial correlations to be learned while preventing contamination by lower-level errors (Section~\ref{sec:HeBzPES}).  The resulting potential exhibits physically correct short- and long-range behavior, even in sparsely sampled regions, and corrects artifacts present in both standard single-fidelity Gaussian process models and empirical approaches utilizing damping functions. The multifidelity \ac{hebz} interaction potential has substantially improved predictive accuracy, and robustness across the domain relevant for many-body simulations.  When applied to grand canonical path integral Monte Carlo simulations of helium adsorption on benzene, the new potential predicts qualitative differences relative to empirical models (Section \ref{sec:pimc}), underscoring the importance of accurately resolving interaction anisotropy and depth in studies of helium–PAH complexes.

\section{Computational Details}

\subsection{Coupled-Cluster Calculations}
 \ac{ccsd_t} calculations with augmented correlation-consistent basis sets (aug-cc-pV$X$Z, $X$ = D, T, Q)\supercite{kendall_electron_1992} were performed for computing the interaction energy of the \ac{hebz} system. All calculations were performed with the frozen core approximation, where the core electrons are not explicitly included in the correlation calculation. The counterpoise (CP)\supercite{BOYS_2002} correction was applied to mitigate basis set superposition error (BSSE), ensuring the reliability of the interaction energies. The reported counterpoise corrected interaction energies were computed as:
\begin{align}
    \Delta E_{\text{int}} = E_{\text{He-Bz}} - E_{\text{Bz-gHe}} - E_{\text{He-gBz}}
    \label{eq:interactionE}    
\end{align}
where $E_{\text{He-Bz}}$ is the \ac{cc} energy of the He-benzene supersystem, while $E_{\text{Bz-gHe}}$ is the energy of benzene with ghost He and $E_{\text{He-gBz}}$ is the energy of He atom with ghost, respectively.
For the estimation of the \ac{hf} and \ac{ccsd_t} energies at the \ac{cbs} limit, we have applied the exponential formula (with $\alpha = 1.63$) and  two-point equations, of Helgaker and coworkers:\supercite{helgaker_basis-set_1997, halkier1999basis}  
\begin{eqnarray}
    E_{\text{HF/CBS}} & = & \frac{E_X - \left(E_Y\times e^{-1.63}\right)}{1- e^{-1.63}} \\
    E_{\text{corr/CBS}} & = & \frac{X^3E_X - Y^3E_Y}{X^3-Y^3} \label{ccsd_eq:1}
\end{eqnarray}
\noindent where $X$ and $Y = X - 1$ are the cardinal numbers for two basis sets. In the next paragraphs, the extrapolated energies are reported as \ac{ccsd_t}/aug-cc-pV(Y,X)Z.

For estimating the error of higher excitations, we extended the coupled-cluster expansion by considering full triples and perturbative quadruple excitations [CCSDT(Q)] with double-zeta correlation-consistent basis sets (cc-pVDZ). The reference electronic energies for all molecular structures were computed by summation of the energy terms from the following composite scheme:
\begin{align}
E_{\text{CCSDT(Q)}} = E_{\text{HF/aug-cc-pV(T,Q)Z}} + \delta E_{\text{CCSD(T)/aug-cc-pV(T,Q)Z}} + \delta E_{\text{T(Q)/cc-pVDZ}}\label{eq:comp_scheme}    
\end{align}
where, 
\begin{align}
 \delta E_{\text{T(Q)}} = \delta E_{\text{CCSDT(Q)/cc-pVDZ}} -  \delta E_{\text{CCSDT/cc-pVDZ}}
\label{eq:T-pertQ}
\end{align}
while the $\delta E$ notation corresponds to correlation energy terms of the total correlation energy. The CCSDT(Q) calculation was performed only at the equilibrium geometry of the He-benzene system.

All \ac{ccsd_t} calculations were performed with the TURBOMOLE 7.7.1\supercite{ahlrichs_electronic_1989} quantum chemical program package. CCSDT and CCSDT(Q) calculations were performed using the MRCC\supercite{kallay_mrcc_2020} package.

\subsection{Density Functional Theory Calculations}
The performance of commonly used density functionals in describing the interaction strength of the \ac{hebz} system is assessed in this study. The following 13 functionals were considered: the generalized gradient approximation PBE\supercite{perdew_generalized_1996} and BLYP\supercite{becke_density-functional_1988, lee_development_1988} functionals, the meta-GGA TPSS\supercite{tao_climbing_2003} functional, the hybrid functionals PBE0\supercite{perdew_rationale_1996}, B3LYP\supercite{xu_x3lyp_2004}, and BHLYP\supercite{becke_new_1993}, the meta-hybrid functionals TPSSh\supercite{staroverov_comparative_2003}, PW6B95, M06\supercite{Zhao_2008}, and M06-2X\supercite{Zhao_2008}, the double hybrid functional B2-PLYP\supercite{grimme_neese_2007}, and the range-separated hybrid functionals CAM-B3LYP\supercite{yanai_tew_handy_2004} and $\omega$B97X\supercite{lin_li_mao_chai_2012}. The potentials were calculated with each of these functionals, using the Ahlrichs basis sets, def2-SVP, def2-SVPD, def2-TZVPP, def2-TZVPPD, def2-QZVPP, and def2-QZVPPD\supercite{weigend_balanced_2005,hellweg_optimized_2007} with grid size m5. Grimme's D4 empirical dispersion correction was added to account for long-range dispersion effects, which are otherwise missing from most of the functionals.\supercite{dftd4-1, dftd4-2, dftd4-rsh, dftd4-actinides, dftd4-periodic} All \ac{dft} calculations were performed with the TURBOMOLE 7.7.1\supercite{ahlrichs_electronic_1989} quantum chemical program package. For assessing the accuracy of the selected density functionals, we have computed the \ac{mae}, \ac{rmse}, and \ac{maxe}:
\begin{align}
    \text{MAE} = \frac{1}{n}\sum_{i=1}^{n}\abs{y_i-x_i}\\
    \text{RMSE} = \sqrt{\frac{1}{n} \sum_{i=1}^{n} (y_i-x_i)^2} \\
    \text{MAXE} = \max_{i}\abs{y_i-x_i}  
\end{align}
where $n$ is the number of grid points (\textit{vide infra}). The CCSD(T) energies were used as the reference.

\subsection{Energy Decomposition Analysis with \ac{sapt}}
\sloppy
\ac{sapt} is a computational method that can be used to obtain systematically increasing accurate interaction energies of non-covalent interactions between atoms and molecules, and it decomposes the total interaction energy into physically meaningful components.\supercite{jeziorski_perturbation_1994} This energy decomposition analysis offers insights into the fundamental nature of intermolecular interactions, revealing the relative contributions of electrostatics (interactions between permanent charge distributions), exchange (Pauli repulsion due to electron overlap), induction (polarization and charge transfer effects), and dispersion (London forces arising from correlated charge fluctuations). \supercite{jeziorski_perturbation_1994}

In this study, \ac{sapt} calculations were performed to address two aims: energy fragmentation for obtaining insights into the weak interactions between helium and benzene, and for generating accurate energies used for the development of potentials. All SAPT calculations were performed at the \ac{sapt}2+3(ccd)dmp2 level of theory with the  aug-cc-pV\textit{X}Z basis sets (\textit{X} = D, T, Q)\supercite{kendall_electron_1992}. The electronic energies were further extrapolated to the \ac{cbs} using Helgaker's formula \eqref{ccsd_eq:1}. For the energy decomposition analysis, we have used the results from the calculations with the larger aug-cc-pVQZ basis set. All \ac{sapt} calculations were performed with Psi4\supercite{smith_psi4_2020} quantum chemical program package.

\subsection{Path Integral Monte Carlo Simulations}

We employ the grand canonical worm algorithm path integral Monte Carlo (PIMC) method \supercite{Ceperley:1995, Boninsegni:2006, pimcrepo} yielding access to expectation values of local and non-local observables $\mathcal{O}$:
\begin{equation}
    \expval{{\mathcal{O}}}  = \frac{1}{\mathcal{Z}} \Tr[\mathcal{O}\ \mathrm{e}^{-\beta {(H-\mu N)}}]\, ,
\label{eq:thermal_expectation_value}
\end{equation}
where $\beta = 1/{k_{\rm B}}T$ is the inverse temperature, $k_{\rm B}$ the Boltzmann constant, $\mu$ is the chemical potential and $\mathcal{Z} = \Tr \mathrm{e}^{-\beta {(H-\mu N)}}$, with $N$ the particle number operator.  $\Tr$ refers to the standard trace operation.  Here $H$ is the many-body Hamiltonian:
\begin{equation}
    H = -\frac{\hbar^2}{2m_{\rm He}}\sum_{i=1}^N \nabla_i^2 + \sum_{i=1}^N V(\vb{x}_i) + \frac{1}{2}\sum_{i,j} V_{\rm He-He}(\vb{x}_i-\vb{x}_j),
\label{eq:Ham}
\end{equation}
where $N$ helium atoms of mass $m_{\rm He}$ at positions $\vb{x}_i = (x_i,y_i,z_i)$
interact through $V_{\rm He-He}$ which is known to high precision, \supercite{Aziz:1979hs,
Przybytek:2010js, Cencek:2012iz} and $V$ is the He-Bz interaction energy.  This method has been extensively used to study the behavior of $^4$He quantum liquids under confinement at low temperatures \supercite{DelMaestro:2011ll,Kulchytskyy:2013dh,Markic:2015bu,Happacher:2013pd,Nava:2022qo} and can provide access to detailed structural and emergent properties for the geometry under consideration.   We are interested in the number and spatial configuration of particles at fixed chemical potential $\mu$ measured in the PIMC via:
\begin{align}
\rho(\vb{x}) &= \expval{\sum_i \delta\qty(\vb{x}-\vb{x}_i)} \\
\expval{N} &= \int \dd{\vb{x}} \rho(\vb{x})
\end{align}
where $\expval{N}$ is computed in practice from the PIMC average of the instantaneous number of closed particle worldlines ($(3+1)$-dimensional space-imaginary time trajectories) in the system. All simulations were performed with an open source PIMC software \cite{pimcrepo}.

\section{High-Fidelity Reference Data Generation}
\label{sec:cc}

\subsection{Interaction Potentials Computed from Coupled-Cluster Calculations}

Our analysis begins with the one-dimensional potential energy profile obtained by placing the helium atom along the axis passing through the center of mass of the benzene molecule, perpendicular to the plane of the molecule. The interaction energies ($\Delta E$\textsubscript{int}) obtained from \ac{ccsd_t} calculations using different basis sets, as well as extrapolations to the \ac{cbs} limit, are summarized in Table \ref{tab:ccsd_t_tab}. The table also lists the equilibrium geometry $R$\textsubscript{min} between He and the center of mass of benzene, and the \ac{mae} for each basis set with respect to the most accurate \ac{ccsd_t}/aug-cc-pV(T, Q)Z energies. As the size of the basis set increases, the \ac{ccsd_t} computed $\Delta E_{\text{int}}$ energies become more negative, indicating a stronger interaction between the helium atom and the benzene molecule. The \ac{mae} also decreases with a larger basis set, signifying improved accuracy with respect to the \ac{ccsd_t}/aug-cc-pV(T,Q)Z extrapolated reference interaction energy. Results obtained with the  aug-cc-pVDZ basis set show the least negative interaction energy (\SI{-61.0}{\centi\meter^{-1}}) and the highest \ac{mae} (\SI{21.0}{\centi\meter^{-1}}). The interaction energies become more negative when larger basis sets are used (\SI{-81.4}{\centi\meter^{-1}} and \SI{-86.3}{\centi\meter^{-1}} for aug-cc-pVTZ and aug-cc-pVQZ, respectively), with \ac{mae}s of \SI{6.1}{\centi\meter^{-1}} and \SI{2.6}{\centi\meter^{-1}}, respectively. The extrapolated CCSD(T)/aug-cc-p(D,T)Z level yields an interaction energy of \SI{-90.9}{\centi\meter^{-1}} with a significantly reduced \ac{mae} of \SI{0.5}{\centi\meter^{-1}}, demonstrating that both (D,T) and (T,Q) extrapolations to the CBS are converging to the same $\Delta E_{\text{int}}$ within \SI{0.6}{\centi\meter^{-1}}. Our results are in close agreement with the previously reported results of Lee et al. (\SI{-89.59}{\centi\meter^{-1}})\supercite{Lee_2003} and Shirkov (\SI{-89.3}{\centi\meter^{-1}}).\supercite{Shirkov_2024}

Furthermore, we evaluated the contribution of the higher-order full triples and perturbative quadruples excitations. We found that they marginally lower the interaction energy by \SI{-0.59}{\centi\meter^{-1}}. Thus, the estimated $\Delta E_{\text{int}}$ at the approximate CCSDT(Q)/CBS level is \SI{-90.79}{\centi\meter^{-1}}. We conclude this part of our study that the CCSD(T)/\ac{cbs} level of theory provides sufficient accuracy for describing the full \ac{hebz} potential. Accordingly, it was chosen to generate the reference electronic energies of the He–benzene system across a range of molecular conformations. 

\begin{figure}[H]
    \centering
    \includegraphics[width=1\linewidth]{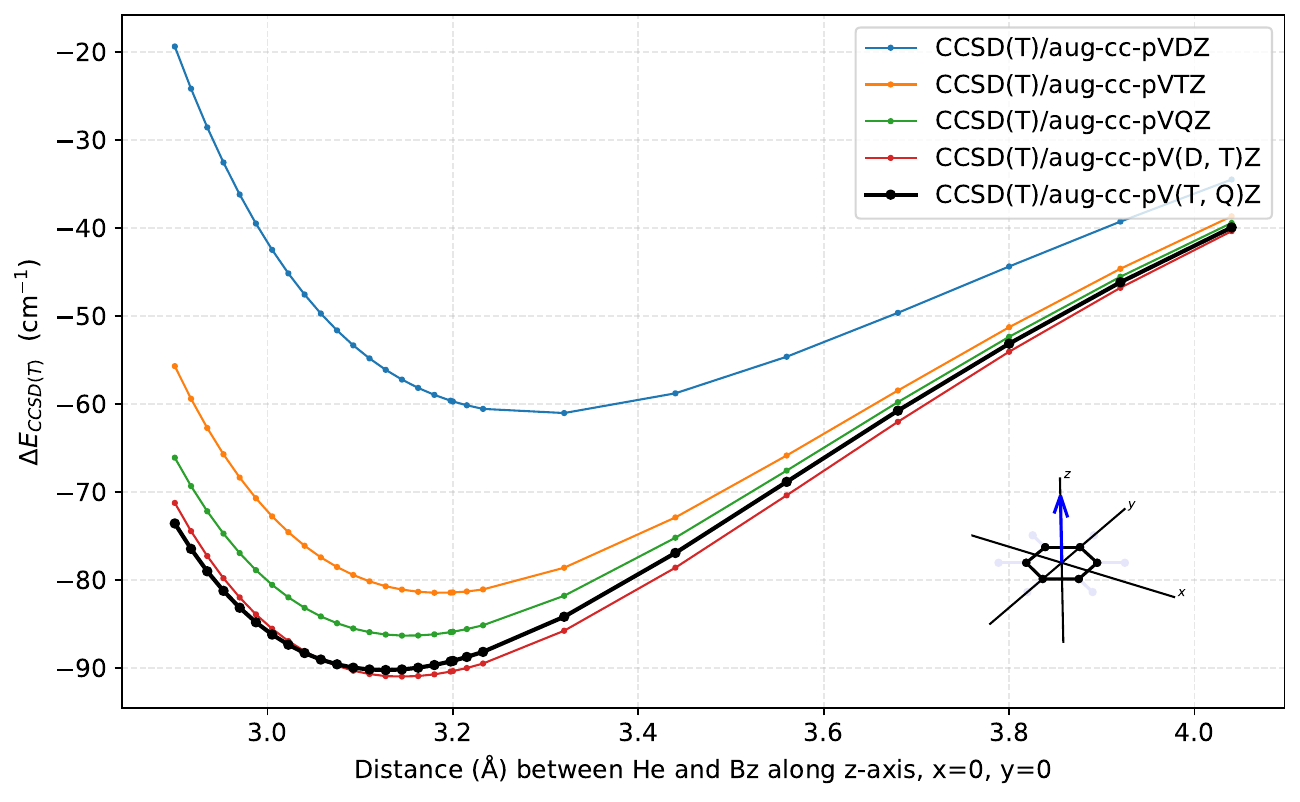}
    \caption{Interaction potential energy curves for a molecular dimer calculated using CCSD(T).  Results are shown for various basis sets of increasing quality: aug-cc-pVDZ, aug-cc-pVTZ, aug-cc-pVQZ, aug-cc-pV(D,T)Z, and aug-cc-pV(T,Q)Z. The aug-cc-pV(T,Q)Z serves as the reference for comparison.  Interaction energies are presented as a function of intermolecular distance. Lines are guides to the eye and the inset shows the 1D cut where the energy is evaluated.}
    \label{fig:ccsd_t_fig}
\end{figure}


\begin{table}[H]
    \centering
    \begin{tabular}{lccccc}
        \hline\hline
        \textbf{Basis set} & \textbf{$R_{\text{min}}$ (\AA)} & \textbf{$\Delta E_{\text{int}}$  (cm\textsuperscript{-1})} & \textbf{MAE} & \textbf{RMSE}& \textbf{MAXE}  \\
        \hline
        aug-cc-pVDZ      & 3.32 & -61.0 & 21.0 & 28.1 & 54.2 \\
        aug-cc-pVTZ      & 3.18 & -81.4 &  6.1 & 8.4 & 17.9 \\
        aug-cc-pVQZ      & 3.15 & -86.3 &  2.6 & 3.5 & 7.5 \\
        aug-cc-pV(D,T)Z  & 3.15 & -90.7 &  0.7 & 0.92 & 2.3 \\
        aug-cc-pV(T,Q)Z  & 3.13 & -90.2 &  0.0 & 0.0 & 0.0 \\
        \hline\hline
    \end{tabular}
    \caption{Statistical errors (MAE, RMSE, and MAXE), in $\si{\centi\meter^{-1}}$, CCSD(T) interaction energies and equilibrium distances between helium and benzene computed with different basis sets and basis set extrapolations.}
    \label{tab:ccsd_t_tab}
\end{table}

\subsection{\ac{dft} Benchmarking}
The performance of various DFT functionals combined with different basis sets was evaluated and benchmarked against the CCSD(T)/aug-cc-pV(T,Q)Z reference data for the He-benzene interaction potential along the 1D cut in the $z$-direction at fixed $x=y=0$. The evaluation metrics used to benchmark the performance of these functionals are the \ac{mae}, \ac{rmse}, and \ac{maxe}. Table \ref{tab:dft_bench} presents the top five best-performing functionals aligning closely with the reference data.

\begin{table}[H]
\centering
\begin{tabular}{lcccc}
 \hline\hline
  \textbf{Functionals} & \textbf{Basis set} & \textbf{MAE} & \textbf{RMSE}& \textbf{MAXE} \\
 \hline
  PBE0      & def2-SVP   &  3.29  &	3.35   &	5.03 \\
  CAM-B3LYP & def2-SVP   &  5.31	&   6.37   &	12.07 \\
  BHLYP     & def2-TZVPP &  7.96  &	9.22   &	16.72 \\
  BHLYP     & def2-SVP   &  10.05	&   10.87  &    16.44 \\
  PBE       & def2-SVP   &  12.14 &	12.24  &    14.51 \\
\hline\hline
\end{tabular}
\caption{Statistical errors (MAE, RMSE, and MAXE), in $\si{\centi\meter^{-1}}$, of DFT benchmark analysis,  for the top 5 functionals for calculating interaction energies in the benzene-methane complex and 2 deviating functionals.  The performance of various functionals is evaluated against the CCSD(T)/aug-cc-pV(T,Q)Z reference data.}
\label{tab:dft_bench}
\end{table}

From this comparison, we conclude that the PBE0 functional combined with the def2-SVP basis set has the highest accuracy among the density functionals and basis sets combinations tested here (lowest \ac{mae} = \SI{3.29}{\centi\meter^{-1}}, \ac{rmse} = \SI{3.35}{\centi\meter^{-1}}, and \ac{maxe} = \SI{5.03}{\centi\meter^{-1}}). This functional is followed by CAM-B3LYP and BHLYP with def2-SVP and def2-TZVPP basis sets, which also show relatively low errors, within 5-10 cm$^{-1}$.

\begin{figure}[t]
    \centering
    \includegraphics[width=1\linewidth]{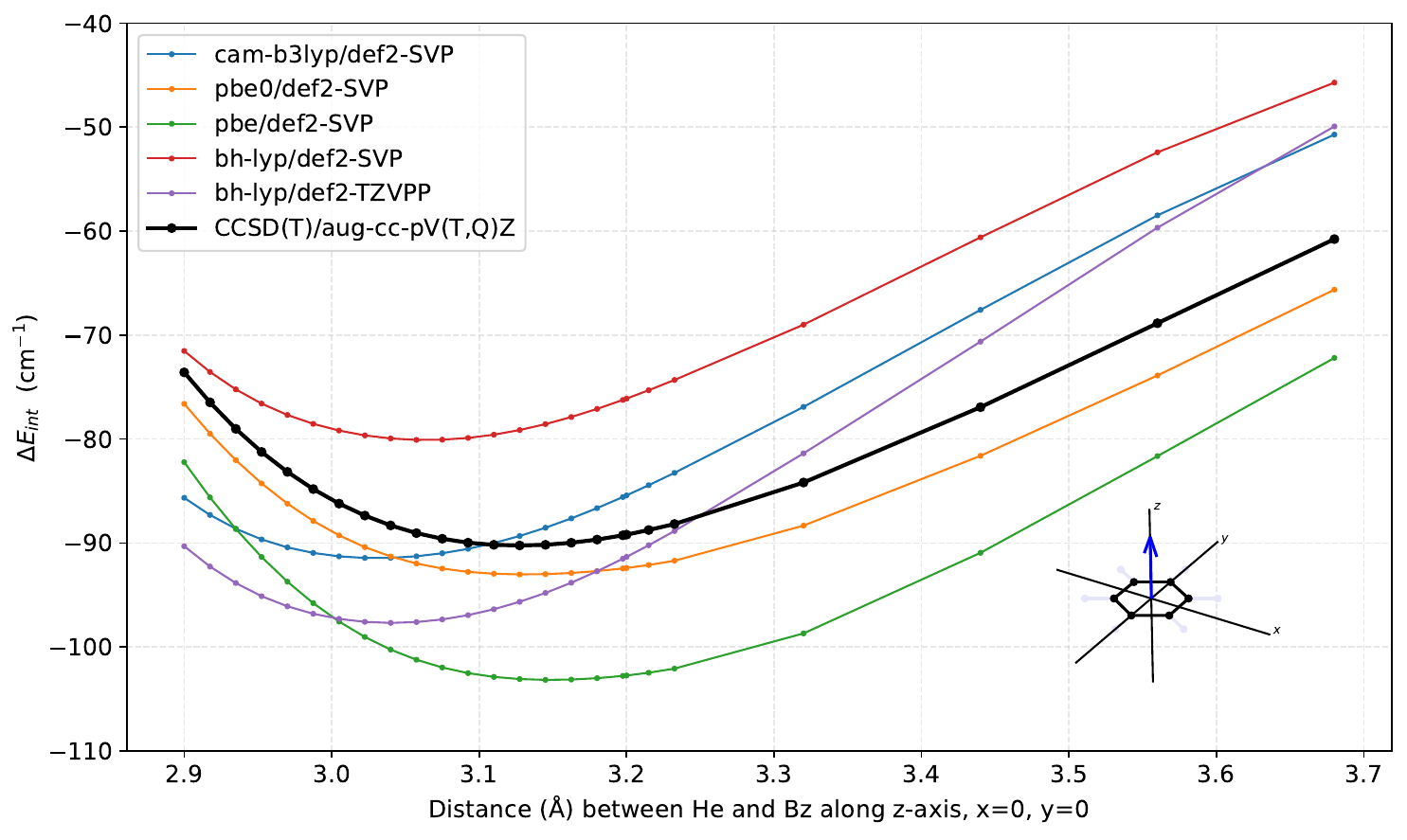}
    \caption{Top 5 best performing DFT functionals bench-marked against CCSD(T)/aug-cc-pV(T,Q)Z (solid black line). Lines are guides to the eye and the inset shows the 1D cut where the energy is evaluated. }
    \label{fig:dft_benchmarking_fig}
\end{figure}

To further test the performance of PBE0, we evaluated different PES cuts with the \ac{ccsd_t}/aug-cc-pV(T,Q)Z reference. These cuts were taken along the $x$-axis (with $y$ = 0) at three distinct $z$-coordinates, representing long-range, equilibrium, and short-range interaction distances.

As shown in Table \ref{tab:pbe0_sapt_bench}, the MAE, RMSE, and MAX Error are relatively negligible at a large separation of $z = \SI{7.0}{\angstrom}$. However, as the interacting systems are brought closer, the error increases significantly. At $z = \SI{3.132}{\angstrom}$, the MAE increases by a factor of six, and at the shortest distance of $z = \SI{2.620}{\angstrom}$, it increases by many folds. This systematic increase in error at smaller $z$ values highlights the difficulty of the PBE0 functional in accurately describing the strong, short-range electron correlations that govern the interaction at close contact.

\subsection{\ac{sapt} Benchmarking}
\begin{figure}[H]
\vspace{-1cm}
  \begin{subfigure}{\textwidth}
    \subcaption{}
    \centering\includegraphics[width=0.75\linewidth,keepaspectratio]{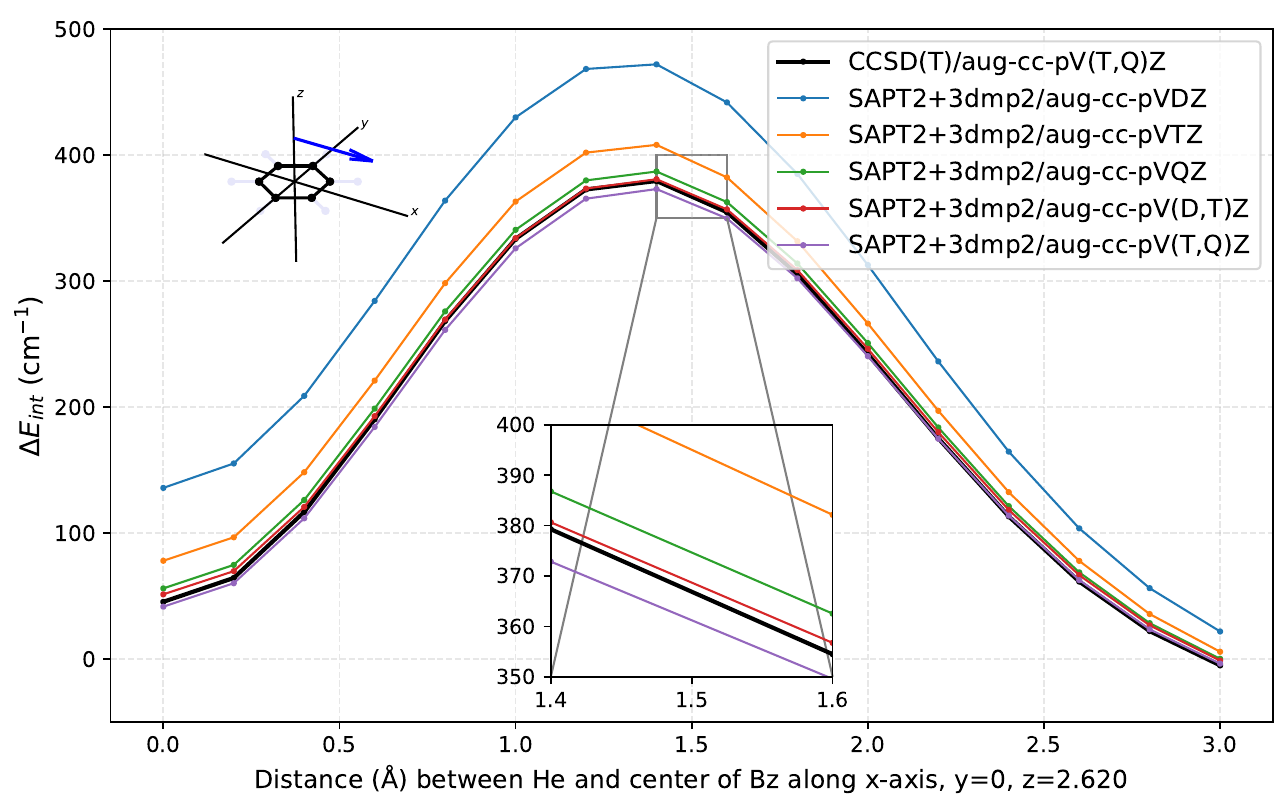}
  \end{subfigure}
 
  \begin{subfigure}{\textwidth}
    \subcaption{}
    \centering\includegraphics[width=0.75\linewidth,keepaspectratio]{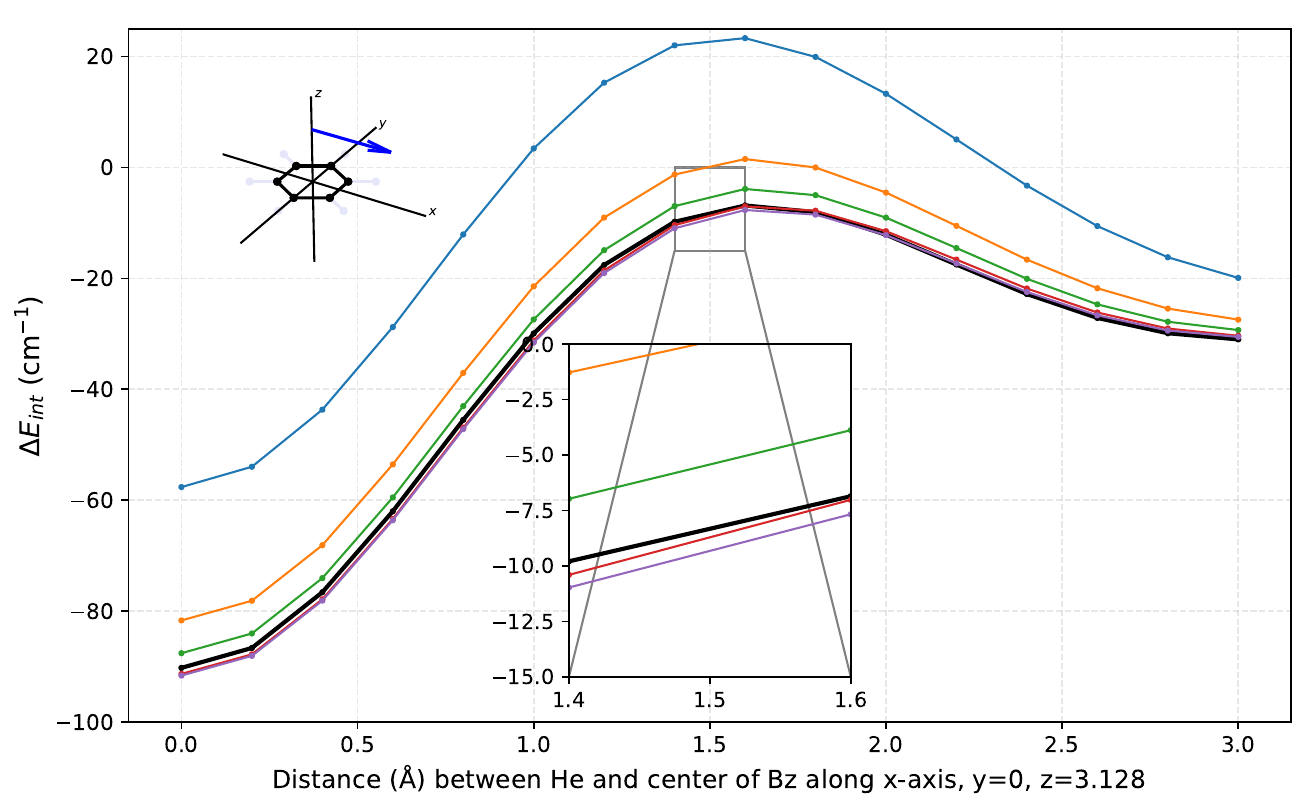}
  \end{subfigure}

  \begin{subfigure}{\textwidth}
    \subcaption{}
    \centering\includegraphics[width=0.75\linewidth,keepaspectratio]{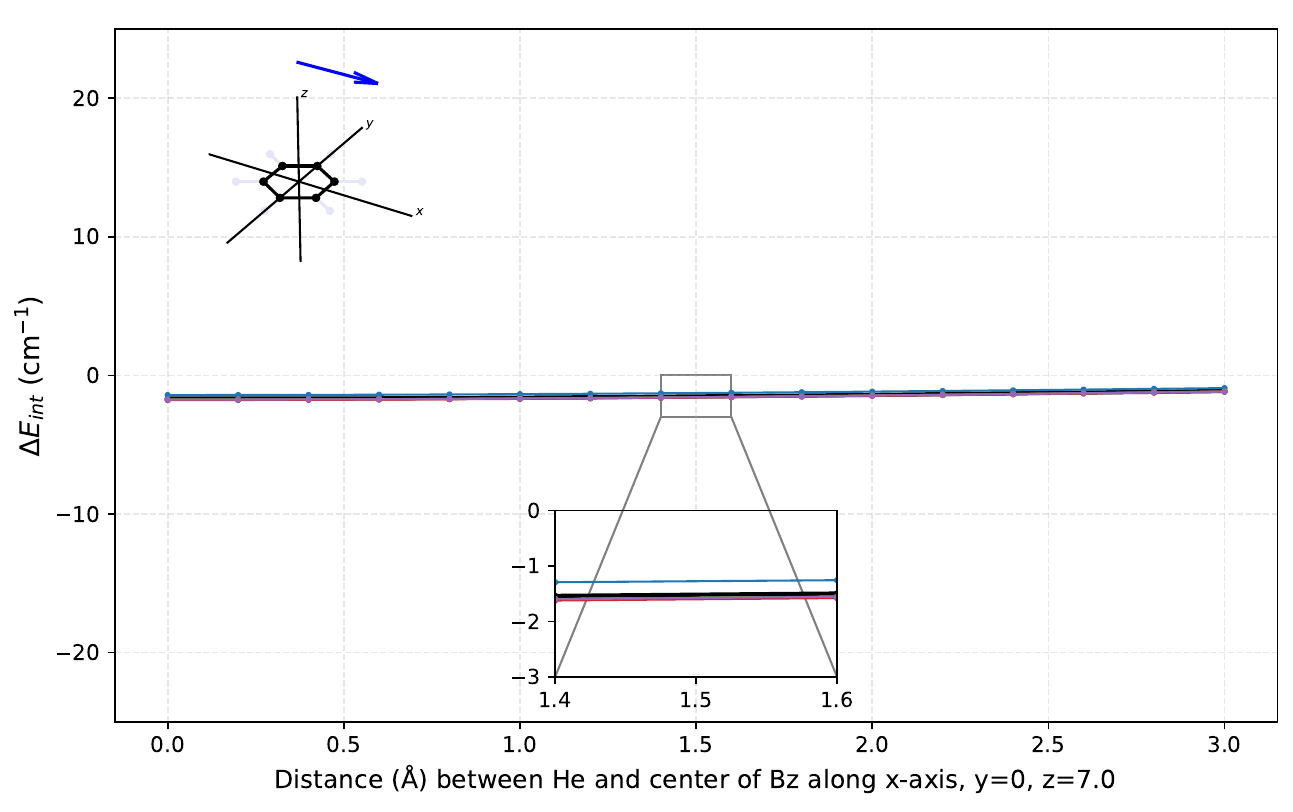}
  \end{subfigure}

  \caption{SAPT2+3dMP2 with different basis sets compared to the reference CCSD(T)/aug-cc-pV(T,Q)Z. Three different potential energy curves where the helium atom moves across the $x$ axis, while $y = 0$ and (a) $z = \SI{2.260}{\angstrom}$, (b) $z = \SI{3.128}{\angstrom}$, (c) $z = \SI{7.000}{\angstrom}$ where the paths are shown in the insets.}
\label{fig:saptfull}
\end{figure}

Given that PBE0 has relatively larger errors at the shorter distances, and in search of a method that balances accuracy and computational efficiency, we evaluated \ac{sapt} performance against the CCSD(T)/\ac{cbs} reference potential along three distinct glancing cuts at fixed $z$-coordinate ($z = \SIlist{2.620;3.128;7.000}{\angstrom}$), representing short-range, equilibrium, and long-range interactions, respectively. As illustrated in Figures \ref{fig:saptfull} (a), (b), and (c), the choice of basis set significantly impacts the accuracy of the interaction energies, especially at shorter intermolecular distances. At long-range ($z = \SI{7.0}{\angstrom}$), calculations with all basis sets (aug-cc-pVDZ, aug-cc-pVTZ, aug-cc-pVQZ) exhibit excellent agreement with the CCSD(T)/aug-cc-pV(T,Q)Z reference. However, at shorter distances, significant deviations emerge. These discrepancies are most pronounced for the smaller aug-cc-pVDZ basis set, whereas the results from the larger aug-cc-pVQZ basis set remain in close agreement with the reference across all distances.

To enhance the accuracy of SAPT calculations using computationally less expensive basis sets, we employed an extrapolation scheme focused on the correlation component of the interaction energy as we discussed previously for \ac{cc}. The total SAPT interaction energy can be partitioned as:

$$\Delta E_{\text{SAPT}} = \Delta E_{\text{HF}} + \Delta E_{\text{corr}}$$
where $\Delta E_{\text{HF}}$ is the Hartree-Fock interaction energy, and $\Delta E_{\text{corr}}$ is the sum of all post-Hartree-Fock corrections (electrostatics, exchange, induction, and dispersion). We applied Helgaker's two-point extrapolation formula to the correlation energy ($\Delta E_{\text{corr}}$), utilizing the results from the aug-ccpVDZ and aug-cc-pVTZ calculations to estimate the CBS limit.

\begin{table}[H]
\centering
\begin{tabular}{lrrrrrr}
\hline \hline
\multirow{2}{*}{\textbf{z-slice (\si{\angstrom})}} & \multicolumn{2}{c}{\textbf{MAE}} & \multicolumn{2}{c}{\textbf{RMSE}} & \multicolumn{2}{c}{\textbf{MAXE}} \\
\cline{2-7}
& \textbf{PBE0} & \textbf{SAPT} & \textbf{PBE0} & \textbf{SAPT} & \textbf{PBE0} & \textbf{SAPT} \\
\hline
\textbf{2.620} & 29.63 & 3.261 & 31.82 & 3.569 & 43.13 & 5.933 \\
\textbf{3.132} & 17.21 & 0.798 & 18.26 & 0.867 & 23.70 & 1.417 \\
\textbf{7.000} & 2.86 & 0.072 & 2.86 & 0.074 & 2.88 & 0.109 \\
\hline \hline
\end{tabular}
\caption{Statistical errors (MAE, RMSE, and MAXE), in $\si{\centi\meter^{-1}}$, for interaction energies ($\Delta E_{\text{int}}$) calculated using the PBE0/Def2-SVP and SAPT/aug-cc-pV(D, T)Z methods. The errors are computed relative to CCSD(T)/aug-cc-pV(T, Q)Z reference values. Data is presented along three 1D glancing cuts with fixed $z$ and $y=0$ corresponding to very close, equilibrium, and large intermolecular distances, respectively.}
\label{tab:pbe0_sapt_bench}
\end{table}


The outcome of this procedure is shown in Figure \ref{fig:saptfull}. The extrapolated potential energy curves demonstrate an improvement in accuracy. Notably, the extrapolated result derived from the \ac{cbs_dt} combination not only achieves a better agreement with the CBS reference but also surpasses the accuracy of the single, more computationally demanding \ac{aqz} calculation. This highlights that a \ac{cbs} extrapolation of the correlation energy is a highly effective strategy for obtaining benchmark-quality results from less costly basis sets.

\subsection{Energy Decomposition Analysis with \ac{sapt}}

The \ac{sapt} energy decomposition analysis provides a comprehensive understanding of the intermolecular interactions in the He-Bz system. The decomposition into exchange, electrostatics, induction, and dispersion energies offers valuable insights into the nature and strength of these interactions. This detailed analysis is crucial for understanding the behavior of molecular systems and can facilitate the development of accurate models for predicting intermolecular interactions. 

\begin{figure}[H]
    \centering
    \includegraphics[width=1\linewidth]{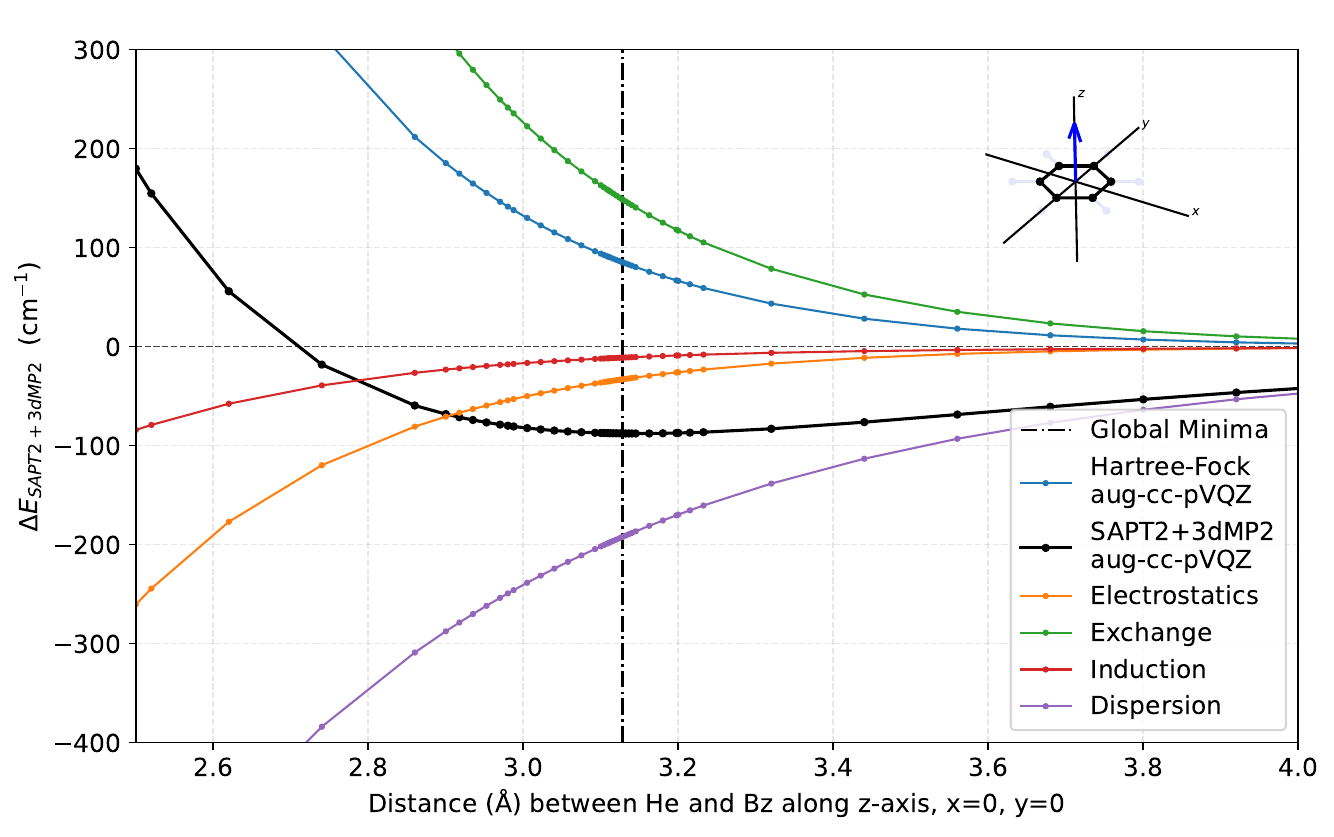}
    \caption{SAPT interaction energy decomposition analysis along the high symmetry trajectory shown in the inset.  Lines are a guide to the eye.}
    \label{fig:sapt_analysis_fig}
\end{figure}

Here, we provide a brief analysis of the equilibrium position \SI{3.14}{\angstrom}. The exchange energy, which represents the Pauli repulsion due to the overlap of the electron clouds of the helium atom and the benzene molecule, is found to be \SI{139.2}{\centi\meter^{-1}}. This positive value indicates a repulsive force between the interacting molecules. The attractive electrostatic energy, which accounts for the Coulombic interactions between the charge distributions of the two molecules, is \SI{-30.9}{\centi\meter^{-1}}. The attractive induction energy, representing the polarization effects where one molecule induces a dipole in the other, is \SI{-10.4}{\centi\meter^{-1}}. The dispersion energy, which arises from correlated fluctuations of the electron density in the two molecules, is \SI{-185.7}{\centi\meter^{-1}}. This component is the largest contributor to the overall interaction, highlighting the significance of dispersion forces in this system. A helium atom has a spherically symmetric electron density and possesses no permanent multipole moments. While benzene is a nonpolar molecule, it has a significant permanent quadrupole moment arising from its electron-rich $\pi$-system. Therefore, the highly polarizable $\pi$-electron cloud of the benzene ring and the electron cloud of the helium atom generate transient, instantaneous dipoles. These transient dipoles interact with each other, resulting in a net attractive force. Combining these components, the total interaction energy is \SI{-87.8}{\centi\meter^{-1}}, indicating a net attractive interaction between helium and benzene. The graph illustrates the variation of these energy components with the He-Bz distance, showing how each term contributes to the interaction as the distance changes. Notably, the exchange energy increases sharply at shorter distances due to stronger repulsion, while the dispersion energy becomes more negative, indicating stronger attraction at longer distances.

\section{Gaussian Process Regression for Mapping the Helium-Benzene Potential Energy Surface}
\label{sec:HeBzPES}
A primary goal of this work is to develop a highly accurate continuous potential energy surface $V(x,y,z)$ over an extended spatial region that can capture helium-benzene interactions suitable for incorporation into many-body simulations.  Due to the computational expense of the coupled-cluster calculations (435 CPU hours per evaluation), we have computed the potential at $n=2595$ discrete points $\vb{x}_i = (x_i,y_i,z_i)$ as seen in Supplementary Material, Fig.~S1. To construct a potential energy surface (PES) from such a discrete data set, a common approach is to perform non-linear least squares fitting to an empirically determined functional form with large numbers of parameters (e.g. Refs.~\cite{Lee_2003,Shirkov_2024}). While such methods often provide reasonable fits, obtaining a low residual between data and model can be challenging, requiring fine-tuning of the measurement grid and fitting regions leading to the possibility of overfitting. \supercite{Schneider2023} Extrapolation and the introduction of artifacts in crossover regions are also a substantial challenge (see Supplementary Material section S3). 

To reduce reliance on any particular analytic model, we instead utilize a Gaussian process (GP) surrogate for the PES. GP regression provides a flexible, nonparametric framework for constructing smooth approximations to functions based on limited training data, \supercite{Rasmussen:2005gp} and they have recently emerged as a powerful tool for modeling potential energy surfaces. \supercite{Uteva2018,Schneider2023, Deringer:2021ky} A GP defines a distribution over functions such that any finite collection of function values follows a multivariate Gaussian distribution, with mean and covariance specified by a kernel function.  The choice of kernel encodes prior assumptions about smoothness, symmetry, and correlation length scales, enabling physically motivated structure to be built into the model. When applied to quantum chemical data, GPs can interpolate the underlying energy landscape with quantified uncertainty, offering both predictive accuracy and error estimates in regions of sparse sampling. This makes them particularly attractive for constructing multidimensional PESs, such as the three-dimensional \ac{hebz} interaction studied here, where the cost of \textit{ab initio} calculations is high and systematic coverage of the full configuration space is prohibitive.

We model the \ac{hebz} interaction potential $V(\vb{x})$ as a realization of a GP defined over all three-dimensional helium coordinates $\vb{x}$. Given a training dataset of positions and corresponding \emph{ab initio} energies,
\begin{equation}
\mathcal{D} = \qty{\vb{X}, \vb{y}} = \qty{(\vb{x}_i, y_i)}_{i=1}^n,\quad \text{where } y_i = V(\vb{x}_i),
\end{equation}
we assume that the potential varies smoothly across space, and that interaction energies $V$ at nearby helium positions $\vb{x}$ are correlated. This correlation structure is captured by a kernel function $k(\vb{x}_i,\vb{x}_j)$, which measures the expected similarity of the interaction energy between two positions $\vb{x}_i$ and $\vb{x}_j$. 

Under this assumption, the collection of energy values is modeled as a multivariate normal distribution,
 $\vb{y} \sim \mathcal{N}\qty(\boldsymbol{m}, \mathsf{K}(\vb{X},\vb{X}))$
%
where the covariance matrix $\mathsf{K}$ has elements
$ \mathsf{K}_{ij} = k(\vb{x}_i,\vb{x}_j) + \sigma_y^2 \delta_{ij}$.
%
Here, $\sigma_y^2$ represents a small noise variance that accounts for numerical uncertainty or residual mismatch between the data and the model, and $\delta_{ij}$ is the Kronecker delta. $m_i = m(\vb{x}_i) = \mathbb{E}\bqty{V(\vb{x}_i)}$ are the components of a (possible) mean vector across the dataset.

To estimate the potential at a new helium position $\vb{x}_{n+1}$ not contained in the dataset, we compute its correlations with all training points through a kernel vector
\begin{equation}
\vb{k}(\vb{x}_{n+1}) =
\qty[k(\vb{x}_1,\vb{x}_{n+1}), \ldots, k(\vb{x}_n,\vb{x}_{n+1})]^{\top}.
\end{equation}
Using this vector and the covariance matrix $\mathsf{K}$, the model interpolates the potential energy at any new coordinate $\vb{x}$ along with a quantitative measure of its uncertainty via:
\begin{align}
    y_{n+1} &= m(\vb{x}_{n+1}) + \vb{k}(\vb{x}_{n+1})^\top \mathsf{K}^{-1} \bqty{\vb{y} - m(\vb{x}_{n+1})} \label{Eq:posteriormean}\\
    \sigma_{n+1}^2 &= k(\vb{x}_{n+1},\vb{x}_{n+1}) - \vb{k}(\vb{x}_{n+1})^\top \mathsf{K}^{-1}\vb{k}(\vb{x}_{n+1})\ . \label{Eq:posteriorconfidenceinterval}
\end{align}

A central modeling choice lies in the kernel function $k(\vb{x},\vb{x}')$ and the mean function $m(\vb{X})$. 
Here we employ a Mat\'ern kernel with stiffness $\nu = 1.5$:
\begin{equation}
k(\vb{x},\vb{x}')
= \sigma^2 \bqty{1 + \sqrt{3}\abs{(\vb{x}-\vb{x}') \oslash \boldsymbol{\ell}}}
\mathrm{e}^{-\sqrt{3}\abs{(\vb{x}-\vb{x}') \oslash \boldsymbol{\ell}}}
\label{eq:MaternThreeHalf}
\end{equation}
where $\vb{x} \oslash \boldsymbol{\ell} = (x_1/\ell_1, x_2/\ell_2, x_3/\ell_3)$ denotes Hadamard division, and the magnitude $\sigma^2$ and per-dimension lengthscale $\boldsymbol{\ell}$ are hyperparameters. This choice was made by manually comparing an exponential kernel (radial basis functions) and Mat\'ern kernels with different values of $\nu$, selecting for accuracy as well as interpolation without oscillating artifacts.  We fix $\sigma_y^2 = 10^{-6}~\si{\centi\meter^{-1}}$ and assume a constant mean function $m$, which can be determined along with the hyperparameters from the training data by maximizing the log-marginal-likelihood,
\begin{equation}
\mathcal{L}(\sigma,\sigma_n,\ell)  = \ln \qty[(2\pi)^{n/2} \abs{\mathsf{K}}^{-1/2} \exp(-\frac{1}{2}\vb{X}^\top \mathsf{K}^{-1} \vb{X})] \, , 
\end{equation}
where $\abs{\mathsf{K}}$ is the determinant.  We perform the optimization using \texttt{botorch} and \texttt{gpytorch} \supercite{botorch:2020,GPyTorch:2018} and obtain an approximation for the PES via Eq.~(\ref{Eq:posteriormean}).  In practice, we employ a truncated dataset $\mathcal{D}^\ast$ where we have removed any points $\vb{x}_i$ corresponding to positions with $y_i > \SI{1000}{cm^{-1}}$ to ensure we are not overfitting the core region of high repulsion where the $^4$He atom is very close to the benzene molecule.  This reduces the original dataset to $n^\ast = 2521$ values.

\begin{figure}[t]
    \centering
    \includegraphics[width=\linewidth]{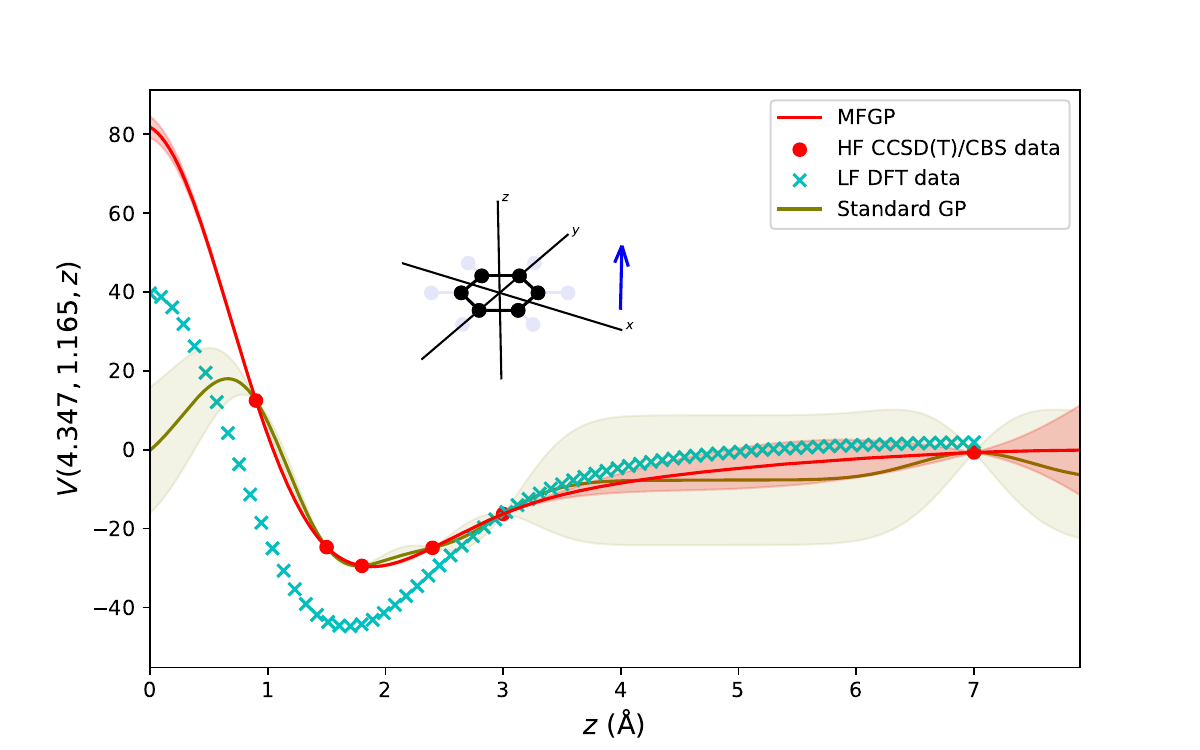}
    \caption{The interaction energy between $^4$He and benzene computed with various approaches along a path $(x=\SI{4.347}{\angstrom}, y=\SI{1.165}{\angstrom},z)$ indicated in the inset.  Solid points are high fidelity coupled cluster calculations (ground truth for fitting) while the crosses are the result of lower fidelity DFT calculations.  Solid lines represent the posterior mean prediction for a standard (green) and multifidelity (red) Gaussian process where details are given in the text.  In both cases the shaded region represents the $1-\sigma$ confidence interval.}
\label{fig:Multifid}
\end{figure}

\begin{figure}[t]
    \centering
    \includegraphics[width=\linewidth]{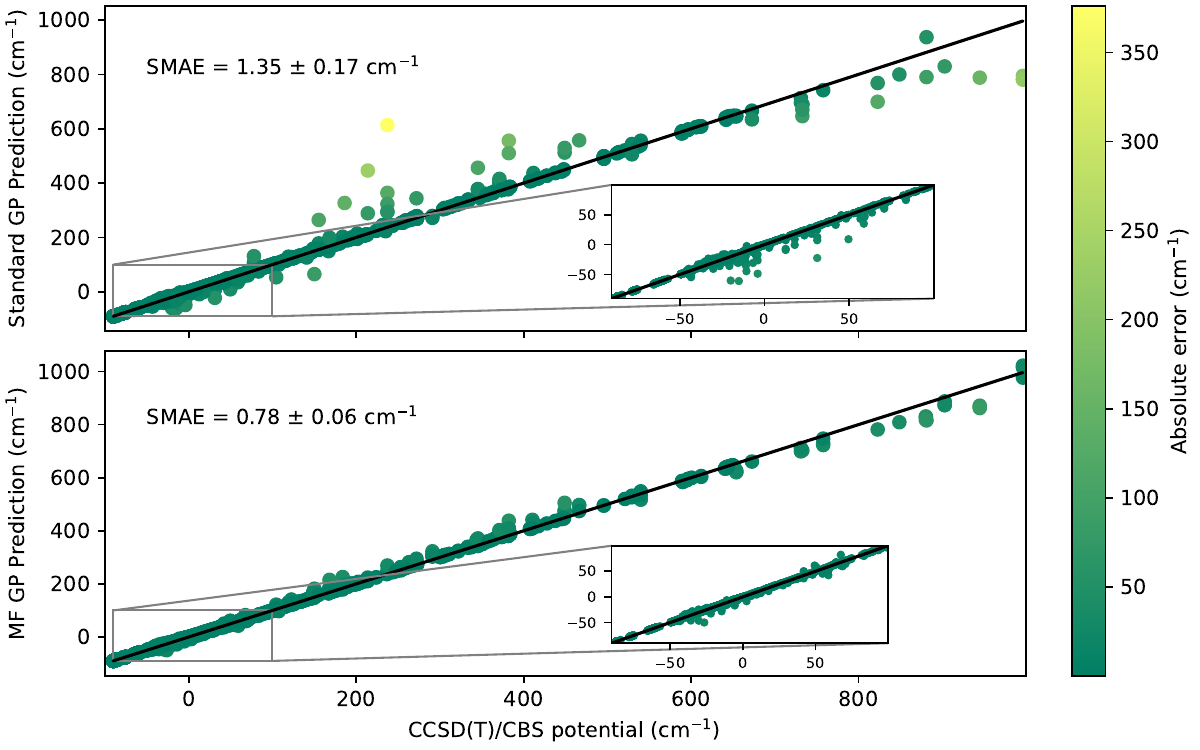}
    \caption{Comparison of the errors between a standard (top) and multi-fidelity
        (bottom) Gaussian process (GP) model predictions for the interaction energy
    between helium and benzene. The reported uncertainty in the sample mean average error is the standard error across $R=8$ realizations. Insets show the low-energy region with more detail and the solid line denotes parity.}
    \label{fig:PotComp}
\end{figure}

To assess the accuracy of our GP model, we would like to measure its prediction fidelity in regions where we do not have any CC data.  This can be quantified by breaking our dataset into a train/test split (we use 80/20) and fitting our GP model $\hat{y}$ over a number $R$ of random realizations of the data. The resulting sampled mean average error (SMAE) is given by: 
\begin{equation}
    \text{SMAE} = \frac{1}{R\, n_{\rm test}} \sum_{r=1}^R\sum_{i=1}^{n_{\rm test}} \abs{\hat{y}^{(r)}(\vb{x}_{r_i}) - y_{r_i}}\, ,
\end{equation}
where $r_i$ is an integer index corresponding to a point in the test set where
we have ground truth data and $\hat{y}^{(r)}$ is the GP model at realization
$r$.  The resulting SMAE for $R=8$ is $\SI{1.35(17)}{\centi\meter^{-1}}$ and the prediction along a path in the $z$-direction with limited CC data is shown as a green line in Fig.~\ref{fig:Multifid}.
While a SMAE approaching a single $\si{\centi\meter^{-1}}$ is certainly low, in order to utilize our GP model for $V$ in quantum many-body simulations we require a potential that is well behaved everywhere in its domain of applicability.  Looking at the GP prediction in Fig~\ref{fig:Multifid} we observe that for the limited CC data we have along this cut, while the model is reasonably accurate near the minimum, there are severe qualitative deviations from the behavior in other regions that are in disagreement with requirements from the physics of the hard-core interaction at small distances (we expect a maximum at $z=0$) and multipole interactions at large distances (we expect $dV/dz < 0$ for $z \gg 1$). This is not surprising as the GP model is ignorant of these physical constraints. The scale of errors can be quantified by investigating a parity plot of GP vs. CC prediction across our dataset as seen in Fig.~\ref{fig:PotComp} (top panel).

Physical deficiencies in the GP model can be mitigated by adopting a multifidelity strategy that augments the training data with low-cost density functional theory calculations, which, while quantitatively inaccurate, capture the correct qualitative behavior of the He–Bz interaction energy.  More generally, multifidelity Gaussian process (MFGP) models provide a Bayesian framework for combining data sets that differ in both accuracy and computational cost \cite{Kennedy2000,Scoggins2023,SabanzaGil:2024bi,Ravi:2024gp}.   By encoding inter-fidelity correlations directly in the GP covariance kernel \cite{Poloczek:2017ni,Mikkola2023}, the MFGP learns how qualitative trends in the low-fidelity data inform the high-fidelity response, while explicitly accounting for their quantitative discrepancies. 

\begin{figure}[t]
    \centering
    \includegraphics[width=\linewidth]{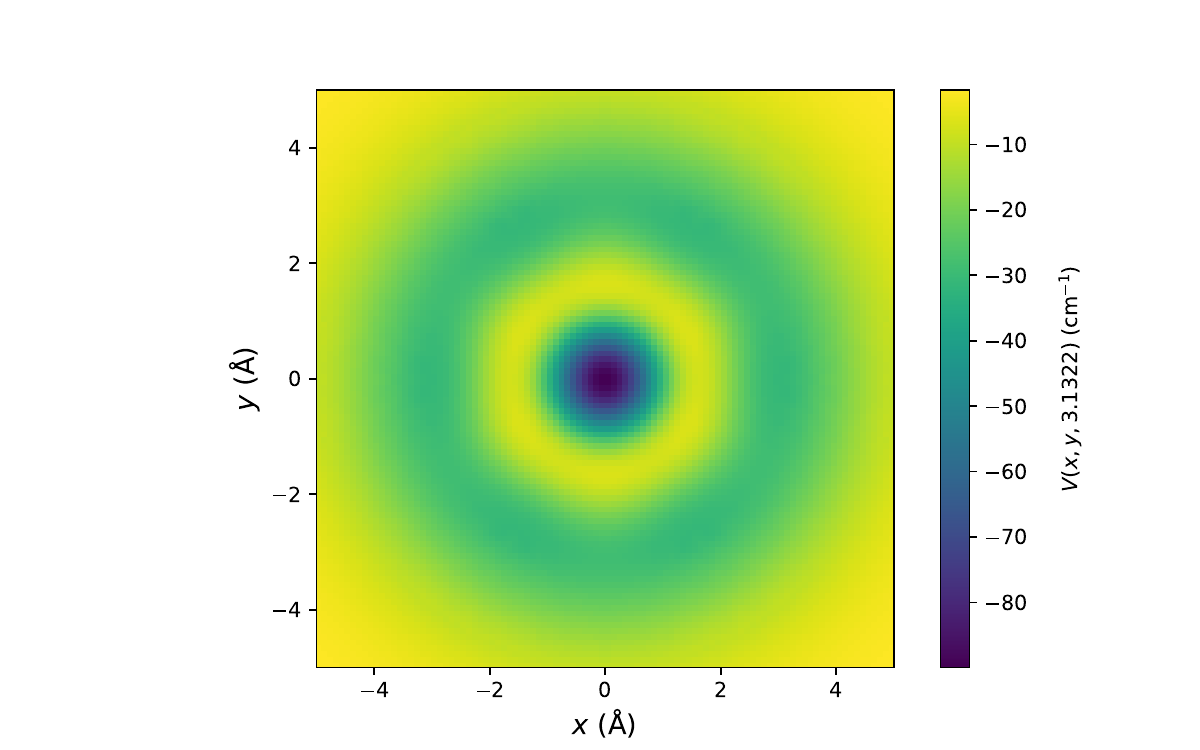}
    \caption{Multifidelity Gaussian process prediction $V(x,y,z^\ast)$ for the interaction potential between helium and benzene at fixed $z^\ast= \text{arg\,min}_z V(0,0,z) = \SI{3.1322}{\angstrom}$.} 
    \label{fig:PotComp2D}
\end{figure}

To proceed we assume that the DFT interaction potential can be described by:
\begin{equation}
    V_{\rm DFT}(\vb{x}) = V(\vb{x}) + \delta(\vb{x}),
    \label{eq:VDFT}
\end{equation}
where $\delta(\vb{x})$ is a bias function at each point in space.  To extend our simple GP model to a multifidelity setting, we augment the input space: $\vb{x} \to  (\vb{x},b)$ where $b \in \bqty{0,1}$ is a binary fidelity variable with $0 \equiv \text{DFT}$ and $1 \equiv \text{CC}$. This yields a modified training dataset: 
\begin{equation}
\tilde{\mathcal{D}} = \Bqty{(\vb{x}_{i}^{(\rm CC)},1,y_i^{(\rm CC)})}_{i=1}^n \cup \Bqty{(\vb{x}_{j}^{(\rm DFT)},0,y_j^{(\rm DFT)})}_{j=1}^q = \Bqty{\tilde{\vb{X}},\tilde{\vb{y}}}\, ,
\end{equation}
where $n \ll q$, $y_j^{(\rm DFT)} = V_{\rm DFT}(\vb{x}_j)$, and in general $\Bqty{\vb{x}_1^{(\rm CC)},\dots,\vb{x}_n^{(\rm CC)}} \neq \Bqty{\vb{x}_{1}^{(\rm DFT)},\dots,\vb{x}_{q}^{(\rm DFT)}}$. Combining with Eq.~(\ref{eq:VDFT}), we model our high CC fidelity data with a MFGP model
\begin{equation}
    \hat{V}(\vb{x}) \sim \mathcal{GP}\qty(m(\vb{x}), k_0(\vb{x},\vb{x}^\prime)), 
\end{equation}
where the systematic bias between DFT and CC is modelled by another GP
\begin{equation}
    \hat{\delta}(\vb{x})  \sim \mathcal{GP}\qty(0,k_1(\vb{x},\vb{x}^\prime))\, ,
\end{equation}
where $k_0,k_1$ are Mat{\'e}rn kernels with $\nu = 2.5$:
\begin{equation}
k(\vb{x},\vb{x}')
= \sigma^2 \bqty{1 + \sqrt{5}\abs{(\vb{x}-\vb{x}') \oslash \boldsymbol{\ell}} + \frac{5}{3}\abs{(\vb{x}-\vb{x}') \oslash \boldsymbol{\ell}}^2}
\mathrm{e}^{-\sqrt{5}\abs{(\vb{x}-\vb{x}') \oslash \boldsymbol{\ell}}}
\label{eq:MaternTwoHalf}
\end{equation}
with identical amplitude $\sigma$, but different lengthscales $\boldsymbol{\ell}_0 \neq \boldsymbol{\ell}_0$. Again this choice was made via a manual search and visual inspection and we identified $\nu=5/2$ as a flexible middle ground between the rougher once differentiable $\nu=3/2$ case in Eq.~(\ref{eq:MaternThreeHalf}) and the unrealistically smooth (infinitely differentiable) exponential kernel.  We have assumed the covariance structure $\text{Cov}(V,V) = k_0$, $\text{Cov}(V,V_{\rm DFT}) = k_0$ and $\text{Cov}(V_{\rm DFT},V_{\rm DFT}) = k_0 + k_1$.  This construction is commonly referred to as the \emph{linear truncated kernel} \cite{Mikkola2023}, as it decomposes the surrogate into a shared latent Gaussian process together with a fidelity-specific bias term that is turned off at the highest fidelity. The resulting prior enforces a triangular information flow: low-fidelity data inform the high-fidelity surrogate through the shared component, while high-fidelity observations do not feed back to modify the low-fidelity model. Within this scheme, the standard GP model is modified as $\tilde{\vb{y}} \sim \mathcal{N}\pqty{m, \tilde{\mathsf{K}}(\tilde{\vb{X}},\tilde{\vb{X}})}$ with 
\begin{equation}
    \tilde{\mathsf{K}}_{i,j} = \sigma^2 \bqty{k_0(\vb{x}_i,\vb{x}_j) + (1-b_i)(1-b_j)k_1(\vb{x}_i,\vb{x}_j)} + \sigma^2_{\tilde{y}} \delta_{i,j}\, ,
\end{equation}
and the MFGP prediction for $V$ is
\begin{equation}
    \hat{V}(\vb{x}) = m(\vb{x}) + \tilde{\vb{k}}(\vb{x})^\top \tilde{\mathsf{K}}^{-1} \bqty{\tilde{\vb{y}} - m(\vb{x})}\, . 
    \label{Eq:MFGPposteriormean}
\end{equation}
For $q=16472$ DFT points, we fix $\sigma^2_{\tilde{y}} = 10^{-6}~\si{\centi\meter^{-1}}$ and the hyperparameters $\sigma,\boldsymbol{\ell}_0,\boldsymbol{\ell}_1$ and a constant mean $m$ are learned from the augmented training data $\tilde{\mathcal{D}}$. The resulting values are given in Table~\ref{tab:params}.
\begin{table}[h!]
    \centering
    \renewcommand{\arraystretch}{1.5}
    \setlength\tabcolsep{5pt}
    \begin{tabular}{@{}ll@{}} 
	\toprule
        \textbf{Parameter} & \textbf{Value} \\
        \midrule
        $\sigma^2$ & $814.69663559~\text{cm}^{-1}$ \\
        $\boldsymbol{\ell}_0$ & $(0.78638807, 2.17270815, 0.77220716)~\si{\angstrom}$ \\
        $\boldsymbol{\ell}_1$ & $(2.03248109, 5.05874827, 1.71032378)~\text{Å}$ \\
        $m$ & $22.0553313~\text{cm}^{-1}$ \\
        \bottomrule
    \end{tabular}
\caption{Learned hyperparameters for the multifidelity linear truncated kernel. High precision is included to aid reproducibility of the model without need for re-training.}
\label{tab:params}
\end{table}
Using the MFGP model for $V$, the MAE is reduced by 40\% to \SI{0.78(0.07)}{\centi\meter^{-1}} over the standard GP. More importantly, physical constraints on the interaction potential that are present in the DFT training data are now fully reflected in the MFGP predicted mean as seen in Fig.~\ref{fig:Multifid}. The improved quality of the model is apparent in the reduced scatter away from prediction parity at all energy scales as shown in the bottom panel of Fig.~\ref{fig:PotComp}.  A two-dimensional slice of the MFGP potential at fixed $z=\SI{3.1423}{\angstrom}$ is shows in Fig.~\ref{fig:PotComp2D}.

To evaluate the interaction potential at large distances from the benzene molecule that are crucial for many-body simulations, we may need to extrapolate outside the region where we have reliable quantum chemical data. For this region, defined as $\mathsf{D} = \left\{ (x,y,z)\in\mathbb{R}^3 \;\middle|\; \sqrt{x^2+y^2} > \SI{5}{\angstrom}\; \text{and}\; z \ge \SI{6.5}{\angstrom}\right\}$, we rely on a standard multipole expansion for the long-range dispersion interaction arising from correlated quantum fluctuations of the charge distributions of the interacting subsystems: \cite{Buckingham:1967os}
%
\begin{equation}
    V_{\rm disp} (\vb*{x}) = -\sum_{n=6,8,10}\sum_{l,m}\frac{C^{l,m}_n}{\abs{\vb*{x}}^n}\Omega_{l,m}(\theta,\phi)\, .
\end{equation}
Here, $\Omega_{l,m}$ are tesseral harmonics and the $D_{6h}$ symmetry of the benzene molecule restricts $(\ell,m) \in \Bqty{(0,0),(2,0),(4,0),(6,0),(6,6),(6,-6)}$.  We use the values of the coefficients $C^{l,m}_n \in \mathbb{R}$ previously reported in the literature from fitting coupled cluster data. \cite{Shirkov_2023,Shirkov_2024,Potentialrepo} The final requirement is that the total potential $V$ is smooth everywhere and that $V_{\rm disp}$ does not contribute at short ranges.  This can be accomplished via an interpolation function: $h(r) = \bqty{1 + \mathrm{e}^{-\gamma(r - r_0)}}^{-1}$ where $\gamma\simeq\SI{16.71}^{\angstrom^{-1}}$ and $r_0 \simeq \SI{5.765}{\angstrom}$ have been previously obtained. \cite{Shirkov_2024}  The final result for the potential is given by the piecewise continuous function:
\begin{equation}
    V(\vb{x}) = \begin{cases} 
        \bqty{1-h(\abs{\vb{x}})}\hat{V}(\vb{x}) + h(\abs{\vb{x}})V_{\rm disp}(\vb{x}) & 
        \; \vb{x} \in \mathsf{D} \\
        \hat{V}(\vb{x}) & \text{otherwise}
   \end{cases}\, .
\end{equation}

\begin{figure}[t]
    \centering
    \includegraphics[width=\linewidth]{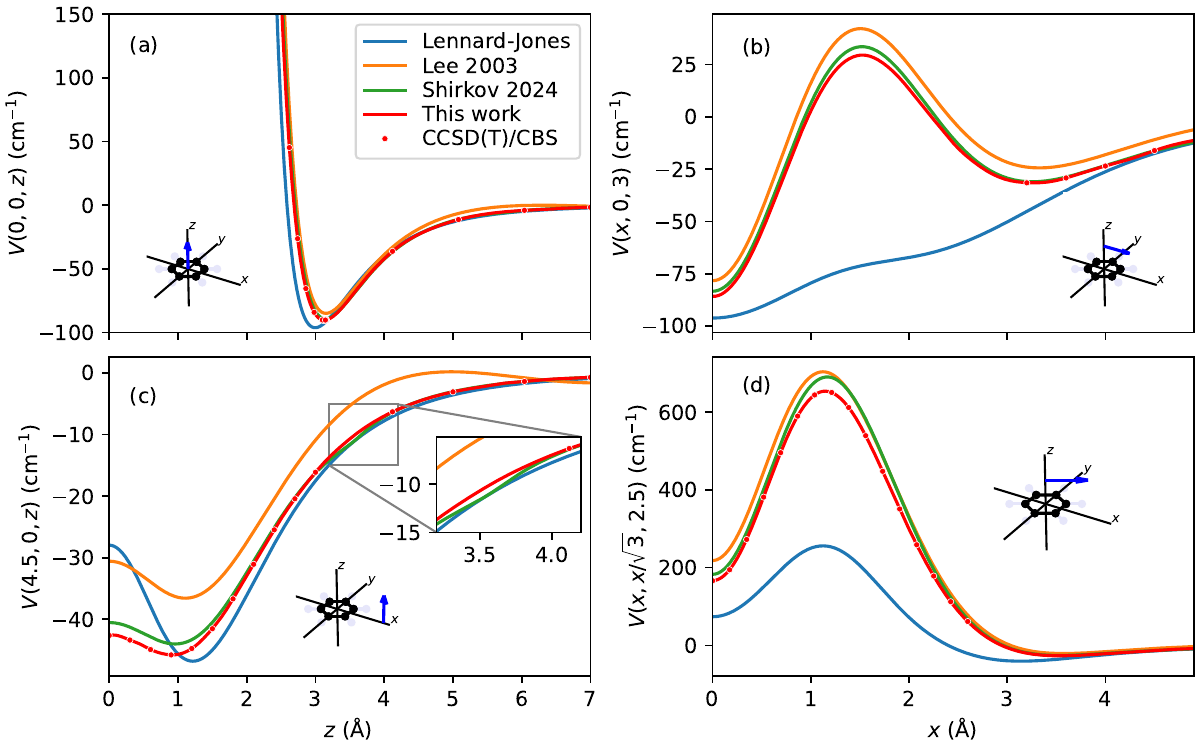}
    \caption{Comparison between this work and different helium-benzene interaction potential functions found in the literature. While in this work we combine coupled cluster data with a multifidelity Gaussian process model, Ref.~\cite{Shirkov_2024} utilizes a different accuracy level of coupled cluster data with an empirical functional form and Ref.~\cite{Lee_2003} uses density functional theory data combined with an empirical fit. Lennard-Jones refers to a pairwise sum of He-atom interactions using $\sigma_{\rm He-C} = \SI{2.98}{\angstrom}, \epsilon_{\rm He-C} =\SI{12.75}{\centi\meter^{-1}}$ and $\sigma_{\rm He-H} = \SI{2.7}{\angstrom}, \epsilon_{\rm He-H} = \SI{8.43}{\centi\meter^{-1}}$ \cite{Boda:2008}. Panels (a) through (d) correspond to the one-dimensional spatial cuts specified in the axis-label and inset. The inset in panel (c) highlights an unphysical feature that can arise when gluing empirical fitting functions via damping (see the Supplementary Material for further analysis).} 
    \label{fig:PotCompCuts}
\end{figure}

Putting everything together (see Fig.~\ref{fig:PotCompCuts}), we compare our MFGP prediction $V$ with \ac{hebz} potentials used or cited in the literature that combine quantum chemical data with empirical functional forms \cite{Lee_2003,Shirkov_2024} along with a simple Lennard-Jones approximation used in many-body simulations \cite{Kwon2001}.
We observe that for the high symmetry path corresponding to $(0,0,z)$ in panel (a), all potentials agree reasonably well, including the simple sum of all Lennard-Jones contributions. For other 1D cuts in 3D space, deviations with Lennard-Jones can be large due to its incorrect assumption of rotationally invariant charge distributions.  The analytically fitted potential obtained from DFT data in \cite{Lee_2003} also shows qualitative agreement in panels (b) and (d) but differs strongly in panel (c) including physically inconsistent oscillations at large distances.  The recent empirical fit to coupled cluster data reported in Ref.~\cite{Shirkov_2024} is very accurate, and agrees with our prediction in most regions of space. However, it does exhibit some small unphysical features  (e.g. a small dip near $z=\SI{3.5}{\angstrom}$ in panel (c)) as a result of gluing different analytical functions together when fitting as confirmed in Section S3 in the supplement. The minima we obtain in panel (c) is a result of extrapolating energies to the complete basis set limit. 


\section{Path Integral Monte Carlo Simulation of $^4$He Nanodroplets}
\label{sec:pimc}

\begin{figure}[h]
    \centering
    \includegraphics[width=\linewidth]{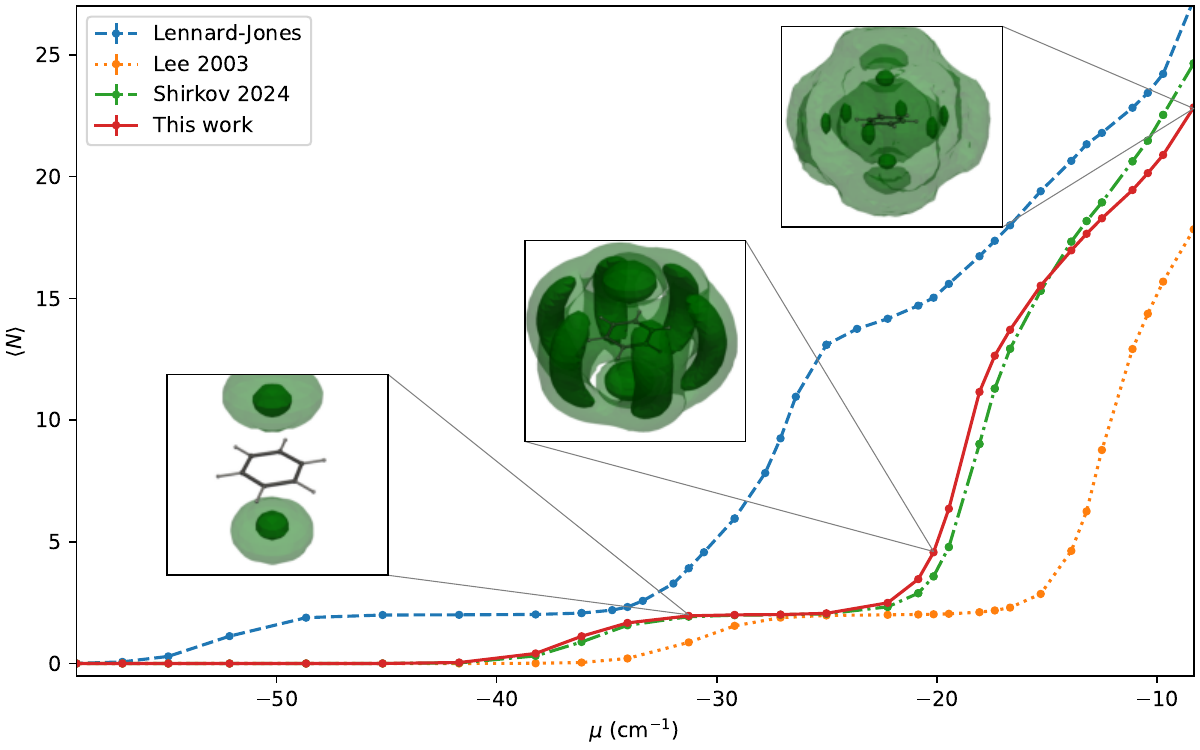}
    \caption{The average number of solvated $^4$He atoms $\expval{N}$ as a function of the chemical potential $\mu$ for different values of the HeBz potential previously described in Fig.~\ref{fig:PotCompCuts} computed via grand canonical path integral quantum Monte Carlo. Lines are guides to the eye. Insets show the particle density at the indicated values of $\mu$.}
    \label{fig:Nmu}
\end{figure}

In this section, we apply the highly accurate potential energy surface (PES) described in Sec.~\ref{sec:HeBzPES} to understand the adsorption and clustering of up to 27 helium atoms surrounding a fixed benzene molecule at low temperatures as a function of the chemical potential $\mu$. We have performed simulations in a cubic box with $L_x=L_y=L_z =\SI{20}{\angstrom}$ at fixed $T = \SI{2.0}{\kelvin}$ below the bulk superfluid transition temperature and measured the average number of particles as a function of chemical potential for four different values of the PES with the results shown in Fig.~\ref{fig:Nmu}. 
As the chemical potential is increased we see a transition from the vacuum state with $\expval{N}=0$ to an incompressible plateau with $\expval{N}=2$ coinciding with a single $^4$He atom strongly adsorbed on either side of the benzene with $z \approx \pm \SI{3.14}{\angstrom}$. The value of $\mu$ where this occurs is strongly dependent on the form of the PES, with the simple Lennard-Jones expression yielding an anomalously strong He-Bz interaction. The Lennard-Jones potential also yields a strong shoulder features near $\mu = \SI{-28}{\centi\meter^{-1}}$ ($\expval{N} \approx 13$) which is absent in the more accurate potentials based on quantum chemical data.   As the chemical potential is further increased, all He-Bz interaction potentials support compressible halo configurations before ultimately the system becomes dominated by the He-He interaction near the bulk saturated vapor pressure chemical potential of helium near $\mu \approx -\SI{6.95}{\centi\meter^{-1}}$ where the simulation box begins to fill towards the low-temperature liquid density \supercite{Donnelly1998}.  It is interesting to note the deviations between the MFGP potential reported in this work and that from Ref.~\cite{Shirkov_2024} which predicts small differences (of order a few $\si{\kelvin}$) in the location and structure of the first solvated cage which consists of $\expval{N} = 10$ atoms.  This can be understood by examining Fig.~\ref{fig:PotCompCuts} (c) which shows that our $V$ has a slightly deeper potential minimum in the plane $z=0$ at $y=0$ and outside the molecule. The observed deviation near $\mu = \SI{-20}{\kelvin}$ is most likely related to the transition to the fully dispersive regime being handled differently here than in previous works \cite{Lee_2003,Shirkov_2024}. 

We can explore the adsorbed and halo configurations more closely by plotting the integrated two-dimensional density:
%
$\rho_{2D}(x,y) = \int_{-L_z/2}^{L_z/2} dz \rho (x,y,z)$
at some interesting values of fixed $\mu$.
\begin{figure}[]
    \centering
    \begin{subfigure}[b]{0.5\textwidth}
        \centering
        \includegraphics[width=\linewidth]{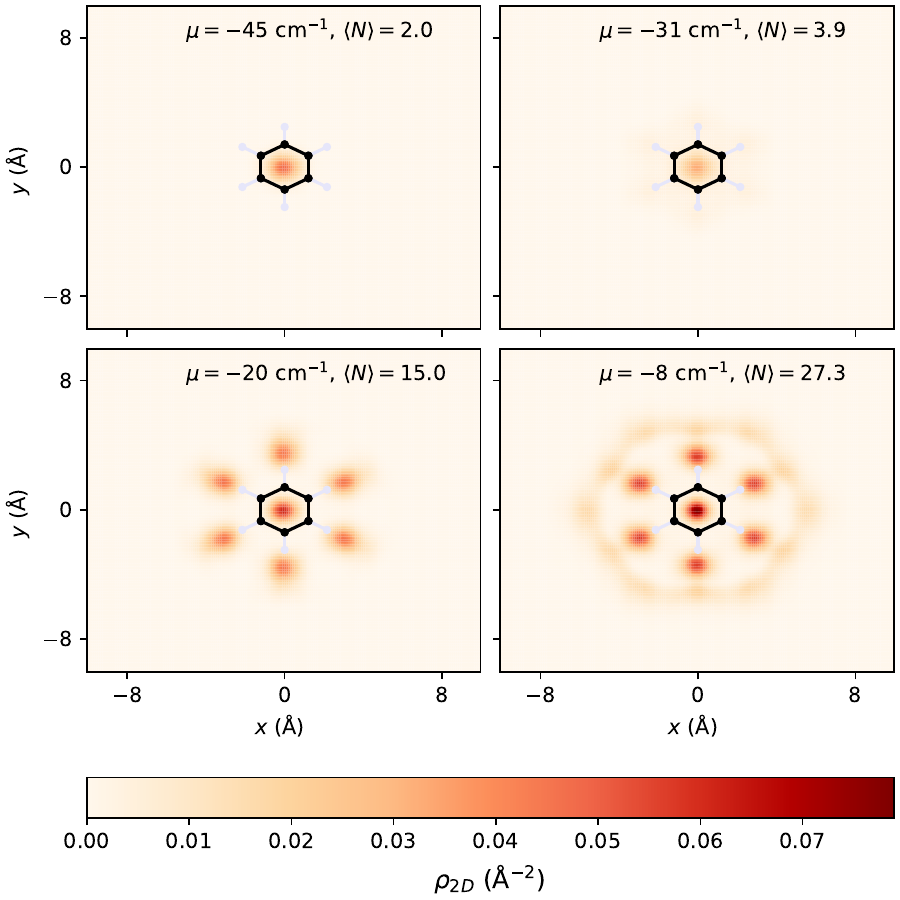}
        \caption{\centering}
    \end{subfigure}%
    \begin{subfigure}[b]{0.5\textwidth}
        \centering
        \includegraphics[width=\linewidth]{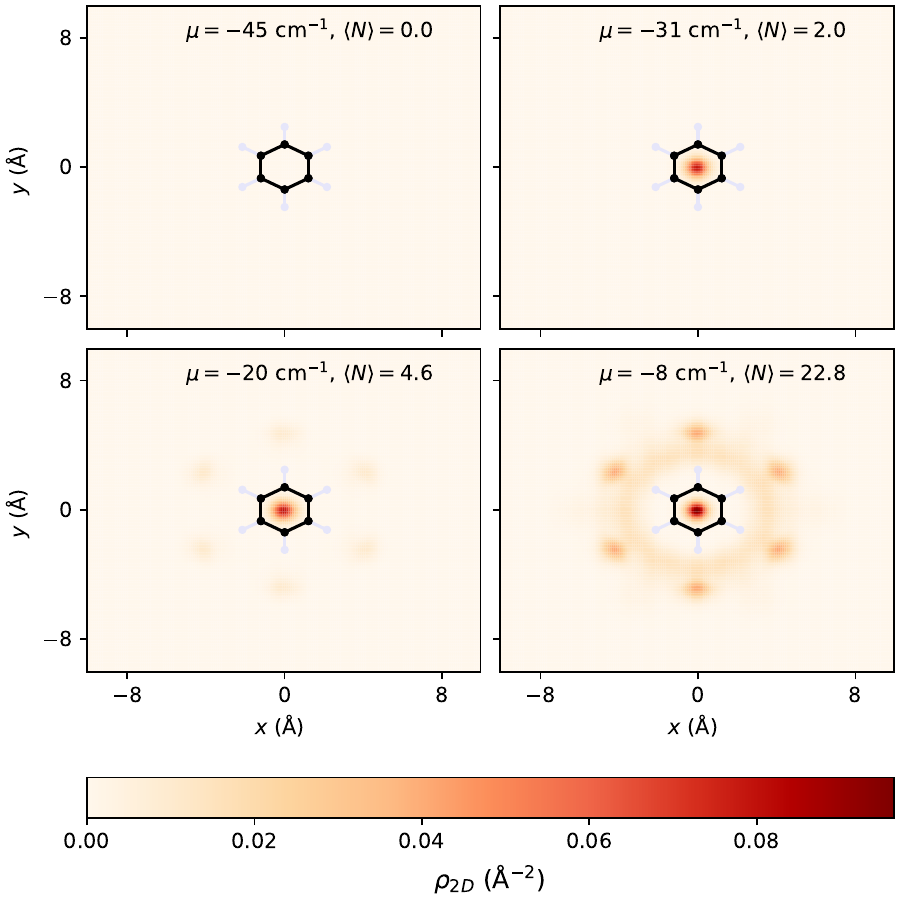}
        \caption{\centering}
    \end{subfigure}
    \caption{Comparison between the evolution of the planar density $\rho_{2D}$ computed with quantum Monte Carlo using (a) Lennard-Jones and (b) a multifidelity gaussian process potential trained on CCSD(T)/CBS data.  The average number of particles $\expval{N}$ for each value of $\mu$ is indicated in the panels. The stronger empirical Lennard-Jones potential is reflected in the apperance of adsorbed $^4$He atoms at $\mu = \SI{-65}{\kelvin}$ in panel (a).}
    \label{fig:density}
\end{figure}
The strong deviations in energy scales and solvated structure between a simple pairwise summation over spherically symmetric Lennard-Jones contributions in panel (a) and the new MFGP potential in panel (b) are striking.  These differences are relevant to simulations and experiments of $^4$He near polycyclic aromatic hydrocarbons including benzene, \cite{Kwon2001,Huang2003} coronene \cite{Cantano2015,Kurzthaler2016,Calvano2017} and larger complexes. \cite{Whitley2005,Cantano2016,kappe2022solvation,Bergmeister2022,Huang2002} Moreover, previous simulations of helium adsorption on two-dimensional carbon surfaces \cite{Pierce1999,Corboz2008,Happacher:2013pd,Gordillo2020,yu_two-dimensional_2021,Moroni:2021lv,Erwin2023} mostly employ a Lennard-Jones (or close analogue) potential and the development of more accurate potentials may explain various levels of disagreement with recent experiments on quasi-2D helium films. \cite{Nakamura2016,Nyki2017,Choi2021,Knapp2025}

    %


\section{Conclusions}


In this work, we have constructed a quantitatively reliable helium–benzene interaction potential by combining high-level electronic structure theory with a physically constrained machine-learning surrogate model. Benchmark CCSD(T) and CCSDT(Q) calculations at the complete basis set limit establish an accurate reference description of the He–Bz interaction across the relevant configuration space, with higher-order correlation effects shown to contribute at the sub-\si{\centi\meter^{-1}} level. Complementary SAPT calculations confirm that the interaction is dominated by dispersion balanced by short-range exchange repulsion, providing a transparent physical decomposition that can be used to interpret both the depth and anisotropy of the interaction potential. We also highlight the inability of DFT to accurately capture the interaction energies for this weakly interacting system.

A central contribution of this study is the construction of a continuous three-dimensional potential energy surface using GP regression trained on our sparse but high accuracy coupled cluster reference dataset. We find that while a standard GP surrogate model yields low mean errors near the potential minimum, it fails to respect known physical constraints in poorly sampled regions, most notably the hard-core repulsion at short distances and the correct asymptotic dispersive behavior at long range. These deficiencies are not numerical artifacts but rather a direct consequence of attempting to interpolate an extremely weak and highly anisotropic interaction using limited high fidelity data.

To address this, we introduced a multifidelity Gaussian process (MFGP) framework that combines our CCSD(T)/CBS reference points with a much larger set of low cost DFT calculations. The resulting surrogate employs a linear truncated kernel that decomposes the interaction into a shared latent process and a fidelity specific bias term, ensuring that our dense DFT data informs the high fidelity potential energy surface through shared spatial correlations, while the coupled cluster data remain unpolluted by DFT-specific errors. This construction substantially reduces both interpolation and extrapolation artifacts, lowers the cross validated mean average error by 40\%, and restores physically correct behavior throughout the domain relevant for many-body simulations. Thus, the MFGP should be viewed not only as enhancing prediction accuracy, but as a mechanism for encoding known physics into the model where ab initio coverage is computationally impractical.

The resulting MFGP potential, augmented by a controlled long-range multipole expansion, provides a smooth and globally well behaved interaction suitable for many-body simulations. When applied to grand canonical path integral Monte Carlo calculations of helium solvation on benzene, the new potential predicts qualitative differences relative to commonly used Lennard-Jones models and earlier empirical fits. In particular, the location and stability of the first solvation shell, as well as the sequence of adsorption plateaus as a function of chemical potential, are sensitive to the detailed anisotropy and depth of the interaction. These differences persist at energy scales relevant for experiments on helium–PAH complexes and underscore the potential consequences of relying on overly simplified, isotropic interaction models as is often done in the field.

More broadly, this work establishes an extensible framework for constructing interaction potentials in weakly bound quantum systems. The combination of SAPT for physical insight, coupled-cluster theory for quantitative accuracy, and multifidelity Gaussian processes for scalable interpolation provides a clear path toward reliable potentials for larger polycyclic aromatic hydrocarbons and, ultimately, 2D periodic materials such as graphene. This will contribute to a deeper understanding of the behavior of helium adsorbed on 2D materials, with potential implications for the design and development of advanced technologies incorporating quantum liquid films, including helium-based qubits.  From a methodological perspective, the MFGP approach introduced here offers a roadmap for future machine learning force fields that retain explicit connections to electronic structure theory rather than replacing it.

\section*{Acknowledgments}
We thank A.~Biswas for insightful discussions.  This research was primarily supported by the National Science Foundation Materials Research Science and Engineering Center program through the UT Knoxville Center for Advanced Materials and Manufacturing (DMR-2309083).

\section*{Code and Data Availability}
All code \supercite{pimcrepo,paperrepo} and data \supercite{datarepo} needed to reproduce
the results of this study are available online. In addition we have a released software to aid in future use of our proposed He-Bz potential. \supercite{Potentialrepo}

\sloppy
\printbibliography

@article{BOYS_2002,
    doi = {10.1080/00268970110088901},
    url = {http://dx.doi.org/10.1080/00268970110088901},
    issn = {1362-3028},
    year = {2002},
    month = {jan},
    pages = {65--73},
    title = {
    The calculation of small molecular interactions by the differences of
    separate total energies. Some procedures with reduced errors
  },
    number = {1},
    volume = {100},
    journal = {Molecular Physics},
    publisher = {Informa UK Limited},
    author = {S. F. BOYS and F. BERNARDI}
}

@article{kim_strain-induced_2024,
    doi = {10.1103/PhysRevB.109.064512},
    url = {http://arxiv.org/abs/2211.07672},
    issn = {2469-9950, 2469-9969},
    note = {arXiv:2211.07672 [cond-mat]},
    year = {2024},
    month = {feb},
    pages = {064512},
    title = {Strain-induced superfluid transition for atoms on graphene},
    number = {6},
    volume = {109},
    journal = {Phys. Rev. B},
    urldate = {2024-07-18},
    author = {Kim, Sang Wook and Elsayed, Mohamed and Nichols, Nathan S. and Lakoba, Taras and Vanegas, Juan and Wexler, Carlos and Kotov, Valeri N. and Del Maestro, Adrian}
}

@article{nichols_adsorption_2016,
    doi = {10.1103/PhysRevB.93.205412},
    url = {http://arxiv.org/abs/1602.04225},
    issn = {2469-9950, 2469-9969},
    note = {arXiv:1602.04225 [cond-mat]},
    year = {2016},
    month = {may},
    pages = {205412},
    title = {Adsorption by design: tuning atom-graphene van der {Waals} interactions via mechanical strain},
    number = {20},
    volume = {93},
    journal = {Physical Review B},
    urldate = {2024-07-18},
    shorttitle = {Adsorption by design},
    author = {Nichols, Nathan S. and Del Maestro, Adrian and Wexler, Carlos and Kotov, Valeri N.}
}

@article{Ravi:2024gp,
 author = {Ravi, Kislaya and Fediukov, Vladyslav and Dietrich, Felix and Neckel, Tobias and Buse, Fabian and Bergmann, Michael and Bungartz, Hans-Joachim},
 doi = {10.48550/arxiv.2404.11965},
 journal = {arXiv:2404.11965},
 title = {{M}ulti-fidelity {G}aussian process surrogate modeling for regression problems in physics},
 url = {https://arxiv.org/abs/2404.11965},
 year = {2024}
}

@inproceedings{Poloczek:2017ni,
 author = {Poloczek, Matthias and Wang, Jialei and Frazier, Peter},
 booktitle = {Advances in Neural Information Processing Systems},
 editor = {I. Guyon and U. Von Luxburg and S. Bengio and H. Wallach and R. Fergus and S. Vishwanathan and R. Garnett},
 pages = {},
 publisher = {Curran Associates, Inc.},
 title = {Multi-Information Source Optimization},
 url = {https://proceedings.neurips.cc/paper_files/paper/2017/file/df1f1d20ee86704251795841e6a9405a-Paper.pdf},
 volume = {30},
 year = {2017}
}

@article{SabanzaGil:2024bi,
 author = {Sabanza-Gil, Víctor and Barbano, Riccardo and Gutiérrez, Daniel Pacheco and Luterbacher, Jeremy S. and Hernández-Lobato, José Miguel and Schwaller, Philippe and Roch, Loïc},
 doi = {10.48550/arxiv.2410.00544},
 journal = {arXiv:2410.00544},
 title = {{B}est {P}ractices for {M}ulti-{F}idelity {B}ayesian {O}ptimization in {M}aterials and {M}olecular {R}esearch},
 url = {https://arxiv.org/abs/2410.00544},
 year = {2024}
}

@article{sengupta_theory_2018,
    doi = {10.1103/PhysRevLett.120.236802},
    url = {http://arxiv.org/abs/1711.09901},
    issn = {0031-9007, 1079-7114},
    note = {arXiv:1711.09901 [cond-mat]},
    year = {2018},
    month = {jun},
    pages = {236802},
    title = {Theory of liquid film growth and wetting instabilities on graphene},
    number = {23},
    volume = {120},
    journal = {Physical Review Letters},
    urldate = {2024-07-18},
    author = {Sengupta, Sanghita and Nichols, Nathan S. and Del Maestro, Adrian and Kotov, Valeri N.}
}

@article{yu_two-dimensional_2021,
    doi = {10.1103/PhysRevB.103.235414},
    url = {https://link.aps.org/doi/10.1103/PhysRevB.103.235414},
    note = {Publisher: American Physical Society},
    year = {2021},
    month = {jun},
    pages = {235414},
    title = {Two-dimensional {Bose}-{Hubbard} model for helium on graphene},
    number = {23},
    volume = {103},
    journal = {Physical Review B},
    urldate = {2024-07-18},
    author = {Yu, Jiangyong and Lauricella, Ethan and Elsayed, Mohamed and Shepherd, Kenneth and Nichols, Nathan S. and Lombardi, Todd and Kim, Sang Wook and Wexler, Carlos and Vanegas, Juan M. and Lakoba, Taras and Kotov, Valeri N. and Del Maestro, Adrian}
}

@article{del_maestro_perspective_2022,
    doi = {10.1002/aelm.202100607},
    url = {http://arxiv.org/abs/2106.07685},
    issn = {2199-160X, 2199-160X},
    note = {arXiv:2106.07685 [cond-mat]},
    year = {2022},
    month = {jan},
    pages = {2100607},
    title = {A {Perspective} on {Collective} {Properties} of {Atoms} on {2D} {Materials}},
    number = {1},
    volume = {8},
    journal = {Advanced Electronic Materials},
    urldate = {2024-07-18},
    author = {Del Maestro, Adrian and Wexler, Carlos and Vanegas, Juan M. and Lakoba, Taras and Kotov, Valeri N.}
}

@article{Choi2021,
  title = {Spatially Modulated Superfluid State in Two-Dimensional He4 Films},
  volume = {127},
  ISSN = {1079-7114},
  url = {http://dx.doi.org/10.1103/physrevlett.127.135301},
  DOI = {10.1103/physrevlett.127.135301},
  number = {13},
  journal = {Physical Review Letters},
  publisher = {American Physical Society (APS)},
  author = {Choi,  Jaewon and Zadorozhko,  Alexey A. and Choi,  Jeakyung and Kim,  Eunseong},
  year = {2021},
  month = sep 
}

@article{helgaker_basis-set_1997,
    doi = {10.1063/1.473863},
    url = {https://doi.org/10.1063/1.473863},
    issn = {0021-9606},
    year = {1997},
    month = {jun},
    pages = {9639--9646},
    title = {Basis-set convergence of correlated calculations on water},
    number = {23},
    volume = {106},
    journal = {The Journal of Chemical Physics},
    urldate = {2024-08-03},
    author = {Helgaker, Trygve and Klopper, Wim and Koch, Henrik and Noga, Jozef}
}

@article{kendall_electron_1992,
    doi = {10.1063/1.462569},
    url = {https://doi.org/10.1063/1.462569},
    issn = {0021-9606},
    year = {1992},
    month = {may},
    pages = {6796--6806},
    title = {Electron affinities of the first‐row atoms revisited. {Systematic} basis sets and wave functions},
    number = {9},
    volume = {96},
    journal = {The Journal of Chemical Physics},
    urldate = {2024-08-03},
    author = {Kendall, Rick A. and Dunning and Harrison, Robert J.}
}

@article{perdew_generalized_1996,
    doi = {10.1103/PhysRevLett.77.3865},
    url = {https://link.aps.org/doi/10.1103/PhysRevLett.77.3865},
    note = {Publisher: American Physical Society},
    year = {1996},
    month = {oct},
    pages = {3865--3868},
    title = {Generalized {Gradient} {Approximation} {Made} {Simple}},
    number = {18},
    volume = {77},
    journal = {Phys. Rev. Lett.},
    urldate = {2024-08-03},
    author = {Perdew, John P. and Burke, Kieron and Ernzerhof, Matthias}
}

@article{xu_x3lyp_2004,
    doi = {10.1073/pnas.0308730100},
    url = {https://www.ncbi.nlm.nih.gov/pmc/articles/PMC374194/},
    issn = {0027-8424},
    pmid = {14981235},
    year = {2004},
    month = {mar},
    pages = {2673--2677},
    pmcid = {PMC374194},
    title = {The {X3LYP} extended density functional for accurate descriptions of nonbond interactions, spin states, and thermochemical properties},
    number = {9},
    volume = {101},
    journal = {Proceedings of the National Academy of Sciences},
    urldate = {2024-08-03},
    author = {Xu, Xin and Goddard, William A.}
}

@article{tao_climbing_2003,
    doi = {10.1103/PhysRevLett.91.146401},
    url = {https://link.aps.org/doi/10.1103/PhysRevLett.91.146401},
    note = {Publisher: American Physical Society},
    year = {2003},
    month = {sep},
    pages = {146401},
    title = {Climbing the {Density} {Functional} {Ladder}: {Nonempirical} {Meta}--{Generalized} {Gradient} {Approximation} {Designed} for {Molecules} and {Solids}},
    number = {14},
    volume = {91},
    journal = {Physical Review Letters},
    urldate = {2024-08-03},
    shorttitle = {Climbing the {Density} {Functional} {Ladder}},
    author = {Tao, Jianmin and Perdew, John P. and Staroverov, Viktor N. and Scuseria, Gustavo E.}
}

@article{becke_density-functional_1988,
    doi = {10.1103/PhysRevA.38.3098},
    url = {https://link.aps.org/doi/10.1103/PhysRevA.38.3098},
    note = {Publisher: American Physical Society},
    year = {1988},
    month = {sep},
    pages = {3098--3100},
    title = {Density-functional exchange-energy approximation with correct asymptotic behavior},
    number = {6},
    volume = {38},
    journal = {Phys. Rev. A},
    urldate = {2024-08-03},
    author = {Becke, A. D.}
}

@article{lee_development_1988,
    doi = {10.1103/physrevb.37.785},
    issn = {0163-1829},
    pmid = {9944570},
    year = {1988},
    month = {jan},
    pages = {785--789},
    title = {Development of the {Colle}-{Salvetti} correlation-energy formula into a functional of the electron density},
    number = {2},
    volume = {37},
    journal = {Phys Rev B Condens Matter},
    language = {eng},
    author = {Lee, C. and Yang, W. and Parr, R. G.}
}

@article{perdew_rationale_1996,
    doi = {10.1063/1.472933},
    url = {https://doi.org/10.1063/1.472933},
    issn = {0021-9606},
    year = {1996},
    month = {dec},
    pages = {9982--9985},
    title = {Rationale for mixing exact exchange with density functional approximations},
    number = {22},
    volume = {105},
    journal = {The Journal of Chemical Physics},
    urldate = {2024-08-03},
    author = {Perdew, John P. and Ernzerhof, Matthias and Burke, Kieron}
}

@article{becke_new_1993,
    doi = {10.1063/1.464304},
    url = {https://doi.org/10.1063/1.464304},
    issn = {0021-9606},
    year = {1993},
    month = {jan},
    pages = {1372--1377},
    title = {A new mixing of {Hartree}–{Fock} and local density‐functional theories},
    number = {2},
    volume = {98},
    journal = {The Journal of Chemical Physics},
    urldate = {2024-08-03},
    author = {Becke, Axel D.}
}

@article{staroverov_comparative_2003,
    doi = {10.1063/1.1626543},
    url = {https://doi.org/10.1063/1.1626543},
    issn = {0021-9606},
    year = {2003},
    month = {dec},
    pages = {12129--12137},
    title = {Comparative assessment of a new nonempirical density functional: {Molecules} and hydrogen-bonded complexes},
    number = {23},
    volume = {119},
    journal = {The Journal of Chemical Physics},
    urldate = {2024-08-03},
    shorttitle = {Comparative assessment of a new nonempirical density functional},
    author = {Staroverov, Viktor N. and Scuseria, Gustavo E. and Tao, Jianmin and Perdew, John P.}
}

@article{weigend_balanced_2005,
    doi = {10.1039/B508541A},
    url = {https://pubs.rsc.org/en/content/articlelanding/2005/cp/b508541a},
    issn = {1463-9084},
    note = {Publisher: The Royal Society of Chemistry},
    year = {2005},
    month = {aug},
    pages = {3297--3305},
    title = {Balanced basis sets of split valence, triple zeta valence and quadruple zeta valence quality for {H} to {Rn}: {Design} and assessment of accuracy},
    number = {18},
    volume = {7},
    journal = {Physical Chemistry Chemical Physics},
    urldate = {2024-08-03},
    language = {en},
    shorttitle = {Balanced basis sets of split valence, triple zeta valence and quadruple zeta valence quality for {H} to {Rn}},
    author = {Weigend, Florian and Ahlrichs, Reinhart}
}

@article{hellweg_optimized_2007,
    doi = {10.1007/s00214-007-0250-5},
    url = {https://doi.org/10.1007/s00214-007-0250-5},
    issn = {1432-2234},
    year = {2007},
    month = {apr},
    pages = {587--597},
    title = {Optimized accurate auxiliary basis sets for {RI}-{MP2} and {RI}-{CC2} calculations for the atoms {Rb} to {Rn}},
    number = {4},
    volume = {117},
    journal = {Theoretical Chemistry Accounts},
    urldate = {2024-08-03},
    language = {en},
    author = {Hellweg, Arnim and Hättig, Christof and Höfener, Sebastian and Klopper, Wim}
}

@article{ahlrichs_electronic_1989,
    doi = {10.1016/0009-2614(89)85118-8},
    url = {https://www.sciencedirect.com/science/article/pii/0009261489851188},
    issn = {0009-2614},
    year = {1989},
    month = {oct},
    pages = {165--169},
    title = {Electronic structure calculations on workstation computers: {The} program system turbomole},
    number = {3},
    volume = {162},
    journal = {Chemical Physics Letters},
    urldate = {2024-08-03},
    shorttitle = {Electronic structure calculations on workstation computers},
    author = {Ahlrichs, Reinhart and Bär, Michael and Häser, Marco and Horn, Hans and Kölmel, Christoph}
}

@article{kallay_mrcc_2020,
    doi = {10.1063/1.5142048},
    issn = {1089-7690},
    pmid = {32087669},
    year = {2020},
    month = {feb},
    pages = {074107},
    title = {The {MRCC} program system: {Accurate} quantum chemistry from water to proteins},
    number = {7},
    volume = {152},
    journal = {J Chem Phys},
    language = {eng},
    shorttitle = {The {MRCC} program system},
    author = {Kállay, Mihály and Nagy, Péter R. and Mester, Dávid and Rolik, Zoltán and Samu, Gyula and Csontos, József and Csóka, József and Szabó, P. Bernát and Gyevi-Nagy, László and Hégely, Bence and Ladjánszki, István and Szegedy, Lóránt and Ladóczki, Bence and Petrov, Klára and Farkas, Máté and Mezei, Pál D. and Ganyecz, Ádám}
}

@article{jeziorski_perturbation_1994,
    doi = {10.1021/cr00031a008},
    url = {https://pubs.acs.org/doi/abs/10.1021/cr00031a008},
    issn = {0009-2665, 1520-6890},
    year = {1994},
    month = {nov},
    pages = {1887--1930},
    title = {Perturbation {Theory} {Approach} to {Intermolecular} {Potential} {Energy} {Surfaces} of van der {Waals} {Complexes}},
    number = {7},
    volume = {94},
    journal = {Chemical Reviews},
    urldate = {2024-08-04},
    language = {en},
    author = {Jeziorski, Bogumil and Moszynski, Robert and Szalewicz, Krzysztof}
}

@article{smith_psi4_2020,
    doi = {10.1063/5.0006002},
    url = {https://doi.org/10.1063/5.0006002},
    issn = {0021-9606},
    year = {2020},
    month = {may},
    pages = {184108},
    title = {{PSI4} 1.4: {Open}-source software for high-throughput quantum chemistry},
    number = {18},
    volume = {152},
    journal = {The Journal of Chemical Physics},
    urldate = {2024-08-04},
    shorttitle = {{PSI4} 1.4},
    author = {Smith, Daniel G. A. and Burns, Lori A. and Simmonett, Andrew C. and Parrish, Robert M. and Schieber, Matthew C. and Galvelis, Raimondas and Kraus, Peter and Kruse, Holger and Di Remigio, Roberto and Alenaizan, Asem and James, Andrew M. and Lehtola, Susi and Misiewicz, Jonathon P. and Scheurer, Maximilian and Shaw, Robert A. and Schriber, Jeffrey B. and Xie, Yi and Glick, Zachary L. and Sirianni, Dominic A. and O’Brien, Joseph Senan and Waldrop, Jonathan M. and Kumar, Ashutosh and Hohenstein, Edward G. and Pritchard, Benjamin P. and Brooks, Bernard R. and Schaefer and Sokolov, Alexander Yu. and Patkowski, Konrad and DePrince and Bozkaya, Uğur and King, Rollin A. and Evangelista, Francesco A. and Turney, Justin M. and Crawford, T. Daniel and Sherrill, C. David}
}

@article{Zhao_2008,
    doi = {10.1007/s00214-007-0401-8},
    url = {http://dx.doi.org/10.1007/s00214-007-0401-8},
    issn = {1432-2234},
    year = {2008},
    month = {jan},
    pages = {525–525},
    title = {The M06 suite of density functionals for main group thermochemistry, thermochemical kinetics, noncovalent interactions, excited states, and transition elements: two new functionals and systematic testing of four M06 functionals and 12 other functionals},
    number = {5–6},
    volume = {119},
    journal = {Theoretical Chemistry Accounts},
    publisher = {Springer Science and Business Media LLC},
    author = {Zhao, Yan and Truhlar, Donald G.}
}

@article{Uteva2018,
  title = {Active learning in Gaussian process interpolation of potential energy surfaces},
  volume = {149},
  url = {http://dx.doi.org/10.1063/1.5051772},
  DOI = {10.1063/1.5051772},
  number = {17},
  journal = {The Journal of Chemical Physics},
  publisher = {AIP Publishing},
  author = {Uteva,  Elena and Graham,  Richard S. and Wilkinson,  Richard D. and Wheatley,  Richard J.},
  year = {2018},
  month = nov 
}

@article{Erwin2023,
  title={Effects of substrate corrugation during helium adsorption on graphene in the grand canonical ensemble},
  author={Erwin, Gage and Del Maestro, Adrian},
  journal={Journal of Low Temperature Physics},
  volume={215},
  number={5},
  pages={525--540},
  year={2024},
  publisher={Springer}
}

@book{Rasmussen:2005gp,
  title={Gaussian Processes for Machine Learning},
  author={Rasmussen, C.E. and Williams, C.K.I.},
  isbn={9780262182539},
  lccn={2005053433},
  url={https://books.google.com/books?id=Tr34DwAAQBAJ},
  year={2005},
  address = {Cambridge, MA},
  publisher={MIT Press}
}

@article{Schneider2023,
  title = {Positioning of grid points for spanning potential energy surfaces—How much effort is really needed?},
  volume = {158},
  ISSN = {1089-7690},
  url = {http://dx.doi.org/10.1063/5.0146020},
  DOI = {10.1063/5.0146020},
  number = {14},
  journal = {The Journal of Chemical Physics},
  publisher = {AIP Publishing},
  author = {Schneider,  Moritz and Born,  Daniel and K\"{a}stner,  Johannes and Rauhut,  Guntram},
  year = {2023},
}

@article{Urade_2022,
    doi = {10.1007/s11837-022-05505-8},
    url = {http://dx.doi.org/10.1007/s11837-022-05505-8},
    issn = {1543-1851},
    year = {2022},
    month = {oct},
    pages = {614–630},
    title = {Graphene Properties, Synthesis and Applications: A Review},
    number = {3},
    volume = {75},
    journal = {JOM},
    publisher = {Springer Science and Business Media LLC},
    author = {Urade, Akanksha R. and Lahiri, Indranil and Suresh, K. S.}
}

@article{Alves_2024,
    doi = {10.1007/s10853-024-10061-4},
    url = {http://dx.doi.org/10.1007/s10853-024-10061-4},
    issn = {1573-4803},
    year = {2024},
    month = {aug},
    title = {Review of scientific literature and standard guidelines for the characterization of graphene-based materials},
    journal = {Journal of Materials Science},
    publisher = {Springer Science and Business Media LLC},
    author = {Alves, Thais and Mota, Wanessa S. and Barros, Cecília and Almeida, Danilo and Komatsu, Daniel and Zielinska, Aleksandra and Cardoso, Juliana C. and Severino, Patrícia and Souto, Eliana B. and Chaud, Marco V.}
}

@article{Guo_2021,
    doi = {10.3390/nano11102539},
    url = {http://dx.doi.org/10.3390/nano11102539},
    issn = {2079-4991},
    year = {2021},
    month = {sep},
    pages = {2539},
    title = {Graphene-Based Films: Fabrication, Interfacial Modification, and Applications},
    number = {10},
    volume = {11},
    journal = {Nanomaterials},
    publisher = {MDPI AG},
    author = {Guo, Sihua and Chen, Jin and Zhang, Yong and Liu, Johan}
}

@article{Sahoo_2024,
    doi = {10.1039/d3ra06904d},
    url = {http://dx.doi.org/10.1039/D3RA06904D},
    issn = {2046-2069},
    year = {2024},
    pages = {1284–1303},
    title = {Recent progress in graphene and its derived hybrid materials for high-performance supercapacitor electrode applications},
    number = {2},
    volume = {14},
    journal = {RSC Advances},
    publisher = {Royal Society of Chemistry (RSC)},
    author = {Sahoo, Prasanta Kumar and Kumar, Niraj and Jena, Anirudha and Mishra, Sujata and Lee, Chuan-Pei and Lee, Seul-Yi and Park, Soo-Jin}
}

@article{Kievsky_2011,
    doi = {10.1007/s00601-011-0226-9},
    url = {http://dx.doi.org/10.1007/s00601-011-0226-9},
    issn = {1432-5411},
    year = {2011},
    month = {mar},
    pages = {259–269},
    title = {The Helium Trimer with Soft-Core Potentials},
    number = {2–4},
    volume = {51},
    journal = {Few-Body Systems},
    publisher = {Springer Science and Business Media LLC},
    author = {Kievsky, A. and Garrido, E. and Romero-Redondo, C. and Barletta, P.}
}

@article{Vranje__2004,
    doi = {10.1016/j.physb.2004.04.069},
    url = {http://dx.doi.org/10.1016/j.physb.2004.04.069},
    issn = {0921-4526},
    year = {2004},
    month = {jun},
    pages = {408–414},
    title = {Helium dimers and trimers within carbon nanotubes},
    number = {1–4},
    volume = {349},
    journal = {Physica B: Condensed Matter},
    publisher = {Elsevier BV},
    author = {Vranješ, Leandra and Antunović, Željko and Kilić, Srećko}
}

@article{Bakr_2013,
    doi = {10.1063/1.4824299},
    url = {http://dx.doi.org/10.1063/1.4824299},
    issn = {1089-7690},
    year = {2013},
    month = {oct},
    title = {Highly accurate potential energy surface for the He–H2 dimer},
    number = {14},
    volume = {139},
    journal = {The Journal of Chemical Physics},
    publisher = {AIP Publishing},
    author = {Bakr, Brandon W. and Smith, Daniel G. A. and Patkowski, Konrad}
}

@article{Borocci_2020,
    doi = {10.1002/jcc.26146},
    url = {http://dx.doi.org/10.1002/jcc.26146},
    issn = {1096-987X},
    year = {2020},
    month = {jan},
    pages = {1000–1011},
    title = {Complexes of helium with neutral molecules: Progress toward a quantitative scale of bonding character},
    number = {10},
    volume = {41},
    journal = {Journal of Computational Chemistry},
    publisher = {Wiley},
    author = {Borocci, Stefano and Grandinetti, Felice and Sanna, Nico and Antoniotti, Paola and Nunzi, Francesca}
}

@article{Donnelly1998,
  title = {The Observed Properties of Liquid Helium at the Saturated Vapor Pressure},
  volume = {27},
  ISSN = {1529-7845},
  url = {http://dx.doi.org/10.1063/1.556028},
  DOI = {10.1063/1.556028},
  number = {6},
  journal = {Journal of Physical and Chemical Reference Data},
  publisher = {AIP Publishing},
  author = {Donnelly,  Russell J. and Barenghi,  Carlo F.},
  year = {1998},
  pages = {1217–1274}
}

@article{Aziz:1979hs,
author = {Aziz, R A and Nain, V P S and Carley, J S and Taylor, W L and McConville, G T},
title = {{An accurate intermolecular potential for helium}},
journal = {The Journal of Chemical Physics},
year = {1979},
volume = {70},
number = {9},
pages = {4330--4342},
publisher = {AIP},
doiOPT = {doi:10.1063/1.438007},
url = {http://dx.doi.org/10.1063/1.438007},
}

@article{pimcrepo,
journal = {{Github Repository}},
title = {{Path Integral Quantum Monte Carlo}},
author = {Adrian {Del Maestro}},
url = {https://code.delmaestro.org},
year = {2024},
doi = {10.5281/zenodo.7271914},
note = {doi:10.5281/zenodo.7271914}
}

@article{Cencek:2012iz,
author = {Cencek, Wojciech and Przybytek, Micha{\l} and Komasa, Jacek and Mehl, James B and Jeziorski, Bogumi{\l} and Szalewicz, Krzysztof},
title = {{Effects of adiabatic, relativistic, and quantum electrodynamics interactions on the pair potential and thermophysical properties of helium}},
journal = {The Journal of Chemical Physics},
year = {2012},
volume = {136},
number = {22},
pages = {224303},
month = jun,
publisher = {AIP Publishing},
doiOPT = {10.1063/1.4712218},
url = {http://scitation.aip.org/content/aip/journal/jcp/136/22/10.1063/1.4712218},
}

@article{Przybytek:2010js,
author = {Przybytek, M and Cencek, W and Komasa, J and {\L}ach, G and Jeziorski, B and Szalewicz, K},
title = {{Relativistic and Quantum Electrodynamics Effects in the Helium Pair Potential}},
journal = {Physical Review Letters},
year = {2010},
volume = {104},
number = {18},
pages = {183003},
month = may,
publisher = {American Physical Society},
doiOPT = {10.1103/PhysRevLett.104.183003},
url = {http://link.aps.org/doi/10.1103/PhysRevLett.104.183003},
}

@article{Cappelletti_2015,
    doi = {10.1002/chem.201406103},
    url = {http://dx.doi.org/10.1002/chem.201406103},
    issn = {1521-3765},
    year = {2015},
    month = {mar},
    pages = {6234–6240},
    title = {Experimental Evidence of Chemical Components in the Bonding of Helium and Neon with Neutral Molecules},
    number = {16},
    volume = {21},
    journal = {Chemistry – A European Journal},
    publisher = {Wiley},
    author = {Cappelletti, David and Bartocci, Alessio and Grandinetti, Felice and Falcinelli, Stefano and Belpassi, Leonardo and Tarantelli, Francesco and Pirani, Fernando}
}

@article{Cappelletti2002,
  title = {Molecular Beam Scattering Experiments on Benzene−Rare Gas Systems:  Probing the Potential Energy Surfaces for the C6H6−He,  −Ne,  and −Ar Dimers},
  volume = {106},
  ISSN = {1520-5215},
  url = {http://dx.doi.org/10.1021/jp0202486},
  DOI = {10.1021/jp0202486},
  number = {45},
  journal = {The Journal of Physical Chemistry A},
  publisher = {American Chemical Society (ACS)},
  author = {Cappelletti,  D. and Bartolomei,  M. and Pirani,  F. and Aquilanti,  V.},
  year = {2002},
  month = sep,
  pages = {10764–10772}
}

@article{Bacanu_2021,
    doi = {10.1063/5.0066817},
    url = {http://dx.doi.org/10.1063/5.0066817},
    issn = {1089-7690},
    year = {2021},
    month = {oct},
    title = {Experimental determination of the interaction potential between a helium atom and the interior surface of a C60 fullerene molecule},
    number = {14},
    volume = {155},
    journal = {The Journal of Chemical Physics},
    publisher = {AIP Publishing},
    author = {Bacanu, George Razvan and Jafari, Tanzeeha and Aouane, Mohamed and Rantaharju, Jyrki and Walkey, Mark and Hoffman, Gabriela and Shugai, Anna and Nagel, Urmas and Jiménez-Ruiz, Monica and Horsewill, Anthony J. and Rols, Stéphane and Rõõm, Toomas and Whitby, Richard J. and Levitt, Malcolm H.}
}

@article{Knapp2025,
  title = {Thermodynamic Evidence for Density Wave Order in a Two Dimensional He Supersolid},
  volume = {134},
  ISSN = {1079-7114},
  url = {http://dx.doi.org/10.1103/physrevlett.134.096002},
  DOI = {10.1103/physrevlett.134.096002},
  number = {9},
  journal = {Physical Review Letters},
  publisher = {American Physical Society (APS)},
  author = {Knapp,  J. and Ny{\'e}ki,  J. and Patel,  H. and Ziouzia,  F. and Cowan,  B.P. and Saunders,  J.},
  year = {2025},
  month = mar 
}

@article{Nakamura2016,
  title = {Possible quantum liquid crystal phases of helium monolayers},
  volume = {94},
  ISSN = {2469-9969},
  url = {http://dx.doi.org/10.1103/physrevb.94.180501},
  DOI = {10.1103/physrevb.94.180501},
  number = {18},
  journal = {Physical Review B},
  publisher = {American Physical Society (APS)},
  author = {Nakamura,  S. and Matsui,  K. and Matsui,  T. and Fukuyama,  Hiroshi},
  year = {2016},
  month = nov 
}

@article{Nyki2017,
  title = {Intertwined superfluid and density wave order in two-dimensional 4He},
  volume = {13},
  ISSN = {1745-2481},
  url = {http://dx.doi.org/10.1038/nphys4023},
  DOI = {10.1038/nphys4023},
  number = {5},
  journal = {Nature Physics},
  publisher = {Springer Science and Business Media LLC},
  author = {Ny\'eki,  J\'an and Phillis,  Anastasia and Ho,  Andrew and Lee,  Derek and Coleman,  Piers and Parpia,  Jeevak and Cowan,  Brian and Saunders,  John},
  year = {2017},
  month = feb,
  pages = {455–459}
}

@article{Moroni:2021lv,
  title = {Localization versus inhomogeneous superfluidity: Submonolayer $^{4}\mathrm{He}$ on fluorographene, hexagonal boron nitride, and graphene},
  author = {Moroni, Saverio and Ancilotto, Francesco and Silvestrelli, Pier Luigi and Reatto, Luciano},
  journal = {Phys. Rev. B},
  volume = {103},
  issue = {17},
  pages = {174514},
  numpages = {11},
  year = {2021},
  month = {May},
  publisher = {American Physical Society},
  doi = {10.1103/PhysRevB.103.174514},
  url = {https://link.aps.org/doi/10.1103/PhysRevB.103.174514}
}

@article{Hayashi2020,
  title = {Quantum Tunneling of a He Atom Above and Below a Benzene Ring},
  volume = {11},
  ISSN = {1948-7185},
  url = {http://dx.doi.org/10.1021/acs.jpclett.0c02879},
  DOI = {10.1021/acs.jpclett.0c02879},
  number = {22},
  journal = {The Journal of Physical Chemistry Letters},
  publisher = {American Chemical Society (ACS)},
  author = {Hayashi,  Masato and Ohshima,  Yasuhiro},
  year = {2020},
  month = nov,
  pages = {9745–9750}
}

@article{Huang2003,
  title = {Localized helium excitations in 4HeN-benzene clusters},
  volume = {67},
  ISSN = {1095-3795},
  url = {http://dx.doi.org/10.1103/physrevb.67.155419},
  DOI = {10.1103/physrevb.67.155419},
  number = {15},
  journal = {Physical Review B},
  publisher = {American Physical Society (APS)},
  author = {Huang,  Patrick and Whaley,  K. Birgitta},
  year = {2003},
  month = apr 
}

@article{Kwon2001,
  title = {Localization of helium at an aromatic molecule in superfluid helium clusters},
  volume = {114},
  ISSN = {1089-7690},
  url = {http://dx.doi.org/10.1063/1.1340567},
  DOI = {10.1063/1.1340567},
  number = {7},
  journal = {The Journal of Chemical Physics},
  publisher = {AIP Publishing},
  author = {Kwon,  Yongkyung and Whaley,  K. Birgitta},
  year = {2001},
  month = feb,
  pages = {3163–3169}
}

@article{Schmied2004,
  title = {UV spectra of benzene isotopomers and dimers in helium nanodroplets},
  volume = {121},
  ISSN = {1089-7690},
  url = {http://dx.doi.org/10.1063/1.1767515},
  DOI = {10.1063/1.1767515},
  number = {6},
  journal = {The Journal of Chemical Physics},
  publisher = {AIP Publishing},
  author = {Schmied,  Roman and \c{C}ar\c{c}abal,  Pierre and Dokter,  Adriaan M. and Lonij,  Vincent P. A. and Lehmann,  Kevin K. and Scoles,  Giacinto},
  year = {2004},
  month = aug,
  pages = {2701–2710}
}

@article{Gao_2020,
    doi = {10.1093/nsr/nwaa064},
    url = {http://dx.doi.org/10.1093/nsr/nwaa064},
    issn = {2053-714X},
    year = {2020},
    month = {apr},
    pages = {1540–1547},
    title = {Coexistence of plastic and partially diffusive phases in a helium-methane compound},
    number = {10},
    volume = {7},
    journal = {National Science Review},
    publisher = {Oxford University Press (OUP)},
    author = {Gao, Hao and Liu, Cong and Hermann, Andreas and Needs, Richard J and Pickard, Chris J and Wang, Hui-Tian and Xing, Dingyu and Sun, Jian}
}

@article{Zunzunegui_Bru_2022,
    doi = {10.1039/d1cp04725f},
    url = {http://dx.doi.org/10.1039/d1cp04725f},
    issn = {1463-9084},
    year = {2022},
    pages = {2004–2014},
    title = {Helium structures around SF5+ and SF6+: novel intermolecular potential and mass spectrometry experiments},
    number = {4},
    volume = {24},
    journal = {Physical Chemistry Chemical Physics},
    publisher = {Royal Society of Chemistry (RSC)},
    author = {Zunzunegui-Bru, Eva and Gruber, Elisabeth and Bergmeister, Stefan and Meyer, Miriam and Zappa, Fabio and Bartolomei, Massimiliano and Pirani, Fernando and Villarreal, Pablo and González-Lezana, Tomás and Scheier, Paul}
}

@article{Rzepa_2010,
    doi = {10.1038/nchem.596},
    url = {http://dx.doi.org/10.1038/nchem.596},
    issn = {1755-4349},
    year = {2010},
    month = {mar},
    pages = {390–393},
    title = {The rational design of helium bonds},
    number = {5},
    volume = {2},
    journal = {Nature Chemistry},
    publisher = {Springer Science and Business Media LLC},
    author = {Rzepa, Henry S.}
}

@article{Shirkov_2023,
    author = {Shirkov, Leonid and Tomza, Michał},
    title = {Long-range interactions of aromatic molecules with alkali-metal and alkaline-earth-metal atoms},
    journal = {The Journal of Chemical Physics},
    volume = {158},
    number = {9},
    pages = {094109},
    year = {2023},
    month = {03},
    issn = {0021-9606},
    doi = {10.1063/5.0135929},
    url = {https://doi.org/10.1063/5.0135929},
    eprint = {https://pubs.aip.org/aip/jcp/article-pdf/doi/10.1063/5.013592989056/094109_1_online.pdf},
}

@article{Shirkov_2024,
    doi = {10.1021/acs.jpca.4c01491},
    url = {http://dx.doi.org/10.1021/acs.jpca.4c01491},
    issn = {1520-5215},
    year = {2024},
    month = {jul},
    pages = {6132–6139},
    title = {Ab Initio Potentials for the Ground S0 and the First Electronically Excited Singlet S1 States of Benzene–Helium with Application to Tunneling Intermolecular Vibrational States},
    number = {30},
    volume = {128},
    journal = {The Journal of Physical Chemistry A},
    publisher = {American Chemical Society (ACS)},
    author = {Shirkov, Leonid}
}

@article{Lee_2003,
    doi = {10.1063/1.1628217},
    url = {http://dx.doi.org/10.1063/1.1628217},
    issn = {1089-7690},
    year = {2003},
    month = {dec},
    pages = {12956–12964},
    title = {Computational and experimental investigation of intermolecular states and forces in the benzene–helium van der Waals complex},
    number = {24},
    volume = {119},
    journal = {The Journal of Chemical Physics},
    publisher = {AIP Publishing},
    author = {Lee, Soohyun and Chung, James S. and Felker, Peter M. and López Cacheiro, Javier and Fernández, Berta and Bondo Pedersen, Thomas and Koch, Henrik}
}

@article{Boninsegni:2006, 
 author = {Boninsegni, M. and Prokof'ev, N.~V. and Svistunov, B.~V.},
 doi = {10.1103/physreve.74.036701},
 journal = {Physical Review E},
 number = {3},
 pages = {036701},
 publisher = {American Physical Society ({APS})},
 title = {{W}orm algorithm and diagrammatic {M}onte {C}arlo: {A} new approach to continuous-space path integral {M}onte {C}arlo simulations},
 url = {https://journals.aps.org/pre/abstract/10.1103/PhysRevE.74.036701},
 volume = {74},
 year = {2006}
}

@article{Ceperley:1995,
author = {Ceperley, D M},
title = {{Path integrals in the theory of condensed helium}},
journal = {Reviews of Modern Physics},
year = {1995},
volume = {67},
number = {2},
pages = {279--355},
month = apr,
doiOPT = {10.1103/RevModPhys.67.279},
url = {http://link.aps.org/doi/10.1103/RevModPhys.67.279},
}

@inproceedings{GPyTorch:2018,
author = {Gardner, Jacob R. and Pleiss, Geoff and Bindel, David and Weinberger, Kilian Q. and Wilson, Andrew Gordon},
title = {GPyTorch: blackbox matrix-matrix Gaussian process inference with GPU acceleration},
year = {2018},
publisher = {Curran Associates Inc.},
address = {Red Hook, NY, USA},
booktitle = {Proceedings of the 32nd International Conference on Neural Information Processing Systems},
pages = {7587–7597},
numpages = {11},
location = {Montr\'{e}al, Canada},
series = {NIPS'18}
}

@inproceedings{botorch:2020,
  title = {{BoTorch: A Framework for Efficient Monte-Carlo Bayesian Optimization}},
  author = {Balandat, Maximilian and Karrer, Brian and Jiang, Daniel R. and Daulton, Samuel and Letham, Benjamin and Wilson, Andrew Gordon and Bakshy, Eytan},
  booktitle = {Advances in Neural Information Processing Systems 33},
  year = 2020,
  url = {http://arxiv.org/abs/1910.06403}
}

@article{Gordillo2020,
  title = {Superfluid and Supersolid Phases of He4 on the Second Layer of Graphite},
  volume = {124},
  ISSN = {1079-7114},
  url = {http://dx.doi.org/10.1103/physrevlett.124.205301},
  DOI = {10.1103/physrevlett.124.205301},
  number = {20},
  journal = {Physical Review Letters},
  publisher = {American Physical Society (APS)},
  author = {Gordillo,  M. C. and Boronat,  J.},
  year = {2020},
  month = may 
}

@article{Pierce1999,
  title = {Path-integral Monte Carlo simulation of the second layer of ${}^{4}\mathrm{He}$ adsorbed on graphite},
  author = {Pierce, Marlon and Manousakis, Efstratios},
  journal = {Phys. Rev. B},
  volume = {59},
  issue = {5},
  pages = {3802--3814},
  numpages = {0},
  year = {1999},
  month = {Feb},
  publisher = {American Physical Society},
  doi = {10.1103/PhysRevB.59.3802},
  url = {https://link.aps.org/doi/10.1103/PhysRevB.59.3802}
}

@article{Corboz2008,
  title = {Phase diagram of $^{4}\text{H}\text{e}$ adsorbed on graphite},
  author = {Corboz, Philippe and Boninsegni, Massimo and Pollet, Lode and Troyer, Matthias},
  journal = {Physical Review B},
  volume = {78},
  issue = {24},
  pages = {245414},
  numpages = {6},
  year = {2008},
  month = {Dec},
  publisher = {American Physical Society},
  doi = {10.1103/PhysRevB.78.245414},
  url = {https://link.aps.org/doi/10.1103/PhysRevB.78.245414}
}

@article{Nichols2016,
  title = {Adsorption by design: Tuning atom-graphene van der Waals interactions via mechanical strain},
  author = {Nichols, Nathan S. and Del Maestro, Adrian and Wexler, Carlos and Kotov, Valeri N.},
  journal = {Physical Review B},
  volume = {93},
  issue = {20},
  pages = {205412},
  numpages = {14},
  year = {2016},
  month = {May},
  publisher = {American Physical Society},
  doi = {10.1103/PhysRevB.93.205412},
  url = {https://link.aps.org/doi/10.1103/PhysRevB.93.205412}
}

@article{Boda:2008,
    author = {Dezs{\"o} Boda and Douglas Henderson},
    title = {The effects of deviations from Lorentz-Berthelot rules on the properties of a simple mixture},
    journal = {Molecular Physics},
    volume = {106},
    number = {20},
    pages = {2367--2370},
    year = {2008},
    publisher = {Taylor \& Francis},
    doi = {10.1080/00268970802471137},
    URL = {https://doi.org/10.1080/00268970802471137},
    eprint = {https://doi.org/10.1080/00268970802471137}
}

@article{dftd4-1,
  author    = {Eike Caldeweyher and Christoph Bannwarth and Stefan Grimme},
  journal   = {The Journal of Chemical Physics},
  title     = {Extension of the D3 dispersion coefficient model},
  year      = {2017},
  month     = {7},
  number    = {3},
  pages     = {034112},
  volume    = {147},
  doi       = {10.1063/1.4993215},
  publisher = {{AIP} Publishing}
}

@article{dftd4-2,
  author    = {Eike Caldeweyher and Sebastian Ehlert and Andreas Hansen and Hagen Neugebauer and Sebastian Spicher and Christoph Bannwarth and Stefan Grimme},
  journal   = {The Journal of Chemical Physics},
  title     = {A generally applicable atomic-charge dependent London dispersion correction},
  year      = {2019},
  month     = {4},
  number    = {15},
  pages     = {154122},
  volume    = {150},
  doi       = {10.1063/1.5090222},
  publisher = {{AIP} Publishing}
}

@article{dftd4-periodic,
  author    = {Eike Caldeweyher and Jan-Michael Mewes and Sebastian Ehlert and Stefan Grimme},
  journal   = {Physical Chemistry Chemical Physics},
  title     = {Extension and evaluation of the D4 London-dispersion model for periodic systems},
  year      = {2020},
  number    = {16},
  pages     = {8499--8512},
  volume    = {22},
  doi       = {10.1039/d0cp00502a},
  publisher = {Royal Society of Chemistry ({RSC})}
}

@article{dftd4-rsh,
  author  = {Friede, Marvin and Ehlert, Sebastian and Grimme, Stefan and Mewes, Jan-Michael},
  title   = {Do Optimally Tuned Range-Separated Hybrid Functionals Require a Reparametrization of the Dispersion Correction? It Depends},
  journal = {Journal of Chemical Theory and Computation},
  volume  = {19},
  number  = {22},
  pages   = {8097-8107},
  year    = {2023},
  doi     = {10.1021/acs.jctc.3c00717},
  note    = {PMID: 37955590},
  url     = {https://doi.org/10.1021/acs.jctc.3c00717}
}

@article{dftd4-actinides,
  author    = {Wittmann, Lukas and Gordiy, Igor and Friede, Marvin and Helmich-Paris, Benjamin and Grimme, Stefan and Hansen, Andreas and Bursch, Markus},
  title     = {Extension of the D3 and D4 London dispersion corrections to the full actinides series},
  journal   = {Physical Chemistry Chemical Physics},
  year      = {2024},
  volume    = {26},
  issue     = {32},
  pages     = {21379-21394},
  publisher = {The Royal Society of Chemistry},
  doi       = {10.1039/D4CP01514B},
  url       = {http://dx.doi.org/10.1039/D4CP01514B},
}

@article{grimme_neese_2007,
  title   = {{Double-hybrid density functional theory for excited electronic states of molecules}},
  volume  = {127},
  url     = {https://pubs.aip.org/aip/jcp/article-abstract/127/15/154116/914655/Double-hybrid-density-functional-theory-for?redirectedFrom=fulltext},
  doi     = {10.1063/1.2772854},
  number  = {15},
  journal = {The Journal of Chemical Physics},
  publisher = {American Institute of Physics},
  author  = {Grimme, Stefan and Neese, Frank},
  year    = {2007},
  month   = {Oct}
}

@article{yanai_tew_handy_2004,
  title   = {{A new hybrid exchange–correlation functional using the Coulomb-attenuating method (CAM-B3LYP)}},
  volume  = {393},
  url     = {https://www.sciencedirect.com/science/article/abs/pii/S0009261404008620?via%3Dihub},
  doi     = {10.1016/j.cplett.2004.06.011},
  number  = {1-3},
  journal = {Chemical Physics Letters},
  publisher = {Elsevier BV},
  author  = {Yanai, Takeshi and Tew, David P and Handy, Nicholas C},
  year    = {2004},
  month   = {Jul},
  pages   = {51–57}
}

@article{lin_li_mao_chai_2012,
  title   = {{Long-Range Corrected Hybrid Density Functionals with Improved Dispersion Corrections}},
  volume  = {9},
  url     = {https://pubs-acs-org.utk.idm.oclc.org/doi/10.1021/ct300715s},
  doi     = {10.1021/ct300715s},
  number  = {1},
  journal = {Journal of Chemical Theory and Computation},
  publisher = {American Chemical Society},
  author  = {Lin, You-Sheng and Li, Guan-De and Mao, Shan-Ping and Chai, Jeng-Da},
  year    = {2012},
  month   = {Nov},
  pages   = {263–272}
}

@article{AlHamdani2021,
  title     = {{Interactions between large molecules pose a puzzle for reference quantum mechanical methods}},
  volume    = {12},
  issn      = {2041-1723},
  url       = {http://dx.doi.org/10.1038/s41467-021-24119-3},
  doi       = {10.1038/s41467-021-24119-3},
  number    = {1},
  journal   = {Nature Communications},
  publisher = {Springer Science and Business Media LLC},
  author    = {Al-Hamdani, Yasmine S. and Nagy, P\'{e}ter R. and Zen, Andrea and Barton, Dennis and K\'{a}llay, Mih\'{a}ly and Brandenburg, Jan Gerit and Tkatchenko, Alexandre},
  year      = {2021},
  month     = jun
}

@article{serhii_tretiakov_nigam_pollice_2025,
title={Studying Noncovalent Interactions in Molecular Systems with Machine Learning}, volume={125}, url={https://pubs-acs-org.utk.idm.oclc.org/doi/10.1021/acs.chemrev.4c00893}, DOI={https://doi.org/10.1021/acs.chemrev.4c00893}, number={12}, journal={Chemical Reviews}, publisher={American Chemical Society}, author={Serhii Tretiakov and Nigam, AkshatKumar and Pollice, Robert}, year={2025}, month={Jun}, pages={5776–5829}
}

@article{al-hamdani_tkatchenko_2019, title={Understanding non-covalent interactions in larger molecular complexes from first principles}, volume={150}, url={https://pubs-aip-org.utk.idm.oclc.org/aip/jcp/article/150/1/010901/152312}, DOI={https://doi.org/10.1063/1.5075487}, number={1}, journal={The Journal of Chemical Physics}, publisher={AIP Publishing}, author={Al-Hamdani, Yasmine S. and Tkatchenko, Alexandre}, year={2019}, month={Jan} }

@article{toennies2013helium,
  title={Helium clusters and droplets: microscopic superfluidity and other quantum effects},
  author={Toennies, J Peter},
  journal={Molecular Physics},
  volume={111},
  number={12-13},
  pages={1879--1891},
  year={2013},
  publisher={Taylor \& Francis}
}

@article{mudrich2014photoionisaton,
  title={Photoionisaton of pure and doped helium nanodroplets},
  author={Mudrich, M and Stienkemeier, F},
  journal={International Reviews in Physical Chemistry},
  volume={33},
  number={3},
  pages={301--339},
  year={2014},
  publisher={Taylor \& Francis}
}

@article{kappe2022solvation,
  title={Solvation of large polycyclic aromatic hydrocarbons in helium: cationic and anionic hexabenzocoronene},
  author={Kappe, Miriam and Calvo, Florent and Sch{\"o}ntag, Johannes and Bettinger, Holger F and Krasnokutski, Serge and Kuhn, Martin and Gruber, Elisabeth and Zappa, Fabio and Scheier, Paul and Echt, Olof},
  journal={Molecules},
  volume={27},
  number={19},
  pages={6764},
  year={2022},
  publisher={MDPI}
}

@article{heidenreich2003permutational,
  title={Permutational symmetry, isotope effects, side crossing, and singlet-triplet splitting in anthracene⋅ He N (N= 1, 2) clusters},
  author={Heidenreich, Andreas and Jortner, Joshua},
  journal={The Journal of Chemical Physics},
  volume={118},
  number={22},
  pages={10101--10119},
  year={2003},
  publisher={American Institute of Physics}
}

@article{felker2003intermolecular,
  title={Intermolecular Hamiltonian for solute--solvent n clusters and application to the (1| 1) isomer of anthracene--He 2},
  author={Felker, Peter M and Neuhauser, Daniel},
  journal={The Journal of Chemical Physics},
  volume={119},
  number={11},
  pages={5558--5569},
  year={2003},
  publisher={American Institute of Physics}
}

@article{heidenreich2001nonrigidity,
  title={Nonrigidity, delocalization, spatial confinement and electronic-vibrational spectroscopy of anthracene--helium clusters},
  author={Heidenreich, Andreas and Even, Uzi and Jortner, Joshua},
  journal={The Journal of Chemical Physics},
  volume={115},
  number={22},
  pages={10175--10185},
  year={2001},
  publisher={American Institute of Physics}
}

@article{xu2007wave,
  title={Wave Function Delocalization and Large-Amplitude Vibrations of Helium on Corrugated Aromatic Microsurfaces: Tetracene⊙ He and Pentacene⊙ He van der Waals Complexes},
  author={Xu, Minzhong and Bači{\'c}, Zlatko},
  journal={The Journal of Physical Chemistry A},
  volume={111},
  number={31},
  pages={7653--7663},
  year={2007},
  publisher={ACS Publications}
}

@article{gibbons2009quantum,
  title={Quantum Dynamics of the Vibrations of Helium Bound to the Nanosurface of a Large Planar Organic Molecule: Phthalocyanine{\textperiodcentered} He van der Waals Complex},
  author={Gibbons, Brittney R and Xu, Minzhong and Bacic, Zlatko},
  journal={The Journal of Physical Chemistry A},
  volume={113},
  number={16},
  pages={3789--3798},
  year={2009},
  publisher={ACS Publications}
}

@article{whitley2009spectral,
  title={Spectral shifts and helium configurations in H4eN--tetracene clusters},
  author={Whitley, Heather D and DuBois, Jonathan L and Whaley, K Birgitta},
  journal={The Journal of Chemical Physics},
  volume={131},
  number={12},
  year={2009},
  publisher={AIP Publishing}
}

@article{whitley2011theoretical,
  title={Theoretical analysis of the anomalous spectral splitting of tetracene in 4He droplets},
  author={Whitley, Heather D and DuBois, Jonathan L and Whaley, K Birgitta},
  journal={The Journal of Physical Chemistry A},
  volume={115},
  number={25},
  pages={7220--7233},
  year={2011},
  publisher={ACS Publications}
}

@article{tao1992mo,
  title={Mo/ller--Plesset perturbation investigation of the He2 potential and the role of midbond basis functions},
  author={Tao, Fu-Ming and Pan, Yuh-Kang},
  journal={The Journal of Chemical Physics},
  volume={97},
  number={7},
  pages={4989--4995},
  year={1992},
  publisher={American Institute of Physics}
}

@article{cybulski1999ground,
  title={Ground state potential energy curves for He 2, Ne 2, Ar 2, He--Ne, He--Ar, and Ne--Ar: a coupled-cluster study},
  author={Cybulski, S{\l}awomir M and Toczy{\l}owski, Rafa{\l} R},
  journal={The Journal of Chemical Physics},
  volume={111},
  number={23},
  pages={10520--10528},
  year={1999},
  publisher={American Institute of Physics}
}

@article{halkier1999basis,
  title={Basis-set convergence of the energy in molecular Hartree--Fock calculations},
  author={Halkier, Asger and Helgaker, Trygve and J{\o}rgensen, Poul and Klopper, Wim and Olsen, Jeppe},
  journal={Chemical Physics Letters},
  volume={302},
  number={5-6},
  pages={437--446},
  year={1999},
  publisher={Elsevier}
}

@article{Cantano2015,
    author = {Rodríguez-Cantano, Rocío and Pérez de Tudela, Ricardo and Bartolomei, Massimiliano and Hernández, Marta I. and Campos-Martínez, José and González-Lezana, Tomás and Villarreal, Pablo and Hernández-Rojas, Javier and Bretón, José},
    title = {Coronene molecules in helium clusters: Quantum and classical studies of energies and configurations},
    journal = {The Journal of Chemical Physics},
    volume = {143},
    number = {22},
    pages = {224306},
    year = {2015},
    month = {12},
    doi = {10.1063/1.4936414},
    url = {https://doi.org/10.1063/1.4936414},
    eprint = {https://pubs.aip.org/aip/jcp/article-pdf/doi/10.1063/1.4936414/15507867/224306_1_online.pdf},
}

@inbook{Huang2002,
    author = {P. Huang and Y. Kwon and K. B. Whaley},
    title = {THE FINITE-TEMPERATURE PATH INTEGRAL MONTE CARLO METHOD AND ITS APPLICATION TO SUPERFLUID HELIUM CLUSTERS},
    booktitle = {Microscopic Approaches to Quantum Liquids in Confined Geometries},
    chapter = {},
    pages = {91-128},
    doi = {10.1142/9789812778475_0003},
    URL = {https://www.worldscientific.com/doi/abs/10.1142/9789812778475_0003},
    eprint = {https://www.worldscientific.com/doi/pdf/10.1142/9789812778475_0003},
    year = {2002},
    publisher={World Scientific}
}

@article{Calvano2017,
    title = {Shell completion of helium atoms around the coronene cation},
    journal = {Computational and Theoretical Chemistry},
    volume = {1107},
    pages = {2-6},
    year = {2017},
    note = {Structure prediction of nanoclusters from global optimization techniques: computational strategies},
    issn = {2210-271X},
    doi = {https://doi.org/10.1016/j.comptc.2016.09.027},
    url = {https://www.sciencedirect.com/science/article/pii/S2210271X16303796},
    author = {F. Calvo},
}

@article{Kurzthaler2016,
    author = {Kurzthaler, Thomas and Rasul, Bilal and Kuhn, Martin and Lindinger, Albrecht and Scheier, Paul and Ellis, Andrew M.},
    title = {The adsorption of helium atoms on coronene cations},
    journal = {The Journal of Chemical Physics},
    volume = {145},
    number = {6},
    pages = {064305},
    year = {2016},
    month = {08},
    issn = {0021-9606},
    doi = {10.1063/1.4960611},
    url = {https://doi.org/10.1063/1.4960611},
    eprint = {https://pubs.aip.org/aip/jcp/article-pdf/doi/10.1063/1.4960611/14724913/064305_1_online.pdf},
}

@article{Whitley2005,
    author = {Whitley, Heather D. and Huang, Patrick and Kwon, Yongkyung and Birgitta Whaley, K.},
    title = {Multiple solvation configurations around phthalocyanine in helium droplets},
    journal = {The Journal of Chemical Physics},
    volume = {123},
    number = {5},
    pages = {054307},
    year = {2005},
    month = {08},
    issn = {0021-9606},
    doi = {10.1063/1.1961532},
    url = {https://doi.org/10.1063/1.1961532},
    eprint = {https://pubs.aip.org/aip/jcp/article-pdf/doi/10.1063/1.1961532/16707780/054307_1_online.pdf},
}

@Article{Bergmeister2022,
    AUTHOR = {Bergmeister, Stefan and Kollotzek, Siegfried and Calvo, Florent and Gruber, Elisabeth and Zappa, Fabio and Scheier, Paul and Echt, Olof},
    TITLE = {Adsorption of Helium and Hydrogen on Triphenylene and 1,3,5-Triphenylbenzene},
    JOURNAL = {Molecules},
    VOLUME = {27},
    YEAR = {2022},
    NUMBER = {15},
    ARTICLE-NUMBER = {4937},
    URL = {https://www.mdpi.com/1420-3049/27/15/4937},
    PubMedID = {35956887},
    ISSN = {1420-3049},
    DOI = {10.3390/molecules27154937}
}

@article{Cantano2016,
    author = {Rocío Rodríguez-Cantano and Tomás González-Lezana and Pablo Villarreal},
    title = {Path integral Monte Carlo investigations on doped helium clusters},
    journal = {International Reviews in Physical Chemistry},
    volume = {35},
    number = {1},
    pages = {37--68},
    year = {2016},
    publisher = {Taylor \& Francis},
    doi = {10.1080/0144235X.2015.1132595},
    URL = {https://doi.org/10.1080/0144235X.2015.1132595},   
    eprint = {https://doi.org/10.1080/0144235X.2015.1132595}
}

@article{Kennedy2000,
 ISSN = {00063444, 14643510},
 URL = {http://www.jstor.org/stable/2673557},
 author = {M. C. Kennedy and A. O'Hagan},
 journal = {Biometrika},
 number = {1},
 pages = {1--13},
 publisher = {[Oxford University Press, Biometrika Trust]},
 title = {Predicting the Output from a Complex Computer Code When Fast Approximations Are Available},
 urldate = {2025-12-01},
 volume = {87},
 year = {2000}
}

@inbook{Scoggins2023,
    author = {James B. Scoggins and Thomas J. Wignall and Tenavi Nakamura-Zimmerer and Karen L. Bibb},
    title = {Multihierarchy Gaussian Process Models for Probabilistic Aerodynamic Databases using Uncertain Nominal and Off-Nominal Configuration Data},
    booktitle = {AIAA SCITECH 2023 Forum},
    chapter = {},
    pages = {},
    doi = {10.2514/6.2023-1185},
    year={2023},
    URL = {https://arc.aiaa.org/doi/abs/10.2514/6.2023-1185},
    eprint = {https://arc.aiaa.org/doi/pdf/10.2514/6.2023-1185},
}

@InProceedings{Mikkola2023,
  title = 	 {Multi-Fidelity Bayesian Optimization with Unreliable Information Sources},
  author =       {Mikkola, Petrus and Martinelli, Julien and Filstroff, Louis and Kaski, Samuel},
  booktitle = 	 {Proceedings of The 26th International Conference on Artificial Intelligence and Statistics},
  pages = 	 {7425--7454},
  year = 	 {2023},
  editor = 	 {Ruiz, Francisco and Dy, Jennifer and van de Meent, Jan-Willem},
  volume = 	 {206},
  series = 	 {Proceedings of Machine Learning Research},
  month = 	 {25--27 Apr},
  publisher =    {PMLR},
  url = 	 {https://proceedings.mlr.press/v206/mikkola23a.html}
}

@article{Kulchytskyy:2013dh,
author = {Kulchytskyy, B and Gervais, G and Del Maestro, A},
title = {{Local superfluidity at the nanoscale}},
journal = {Physical Review B},
year = {2013},
volume = {88},
number = {6},
pages = {064512},
month = aug,
url = {http://link.aps.org/doi/10.1103/PhysRevB.88.064512},
}

@article{DelMaestro:2011ll,
  title = {$^{4}\mathrm{He}$ Luttinger Liquid in Nanopores},
  author = {Del Maestro, Adrian and Boninsegni, Massimo and Affleck, Ian},
  journal = {Physical Review Letters},
  volume = {106},
  issue = {10},
  pages = {105303},
  numpages = {4},
  year = {2011},
  month = {Mar},
  publisher = {American Physical Society},
  doi = {10.1103/PhysRevLett.106.105303},
  url = {http://link.aps.org/doi/10.1103/PhysRevLett.106.105303}
}

@article{Nava:2022qo,
  title = {Quasi-one-dimensional $^{4}\mathrm{He}$ in nanopores},
  author = {Nava, Andrea and Giuliano, Domenico and Nguyen, Phong H. and Boninsegni, Massimo},
  journal = {Physical Review B},
  volume = {105},
  issue = {8},
  pages = {085402},
  numpages = {8},
  year = {2022},
  month = {Feb},
  publisher = {American Physical Society},
  doi = {10.1103/PhysRevB.105.085402},
  url = {https://link.aps.org/doi/10.1103/PhysRevB.105.085402}
}

@article{Markic:2015bu,
  title = {Superfluidity, {BEC}, and dimensions of liquid $^{4}\mathrm{He}$ in nanopores},
  author = {Marki{\'c}, L. {Vranje{\v s}} and Glyde, H. R.},
  journal = {Physical Review B},
  volume = {92},
  issue = {6},
  pages = {064510},
  numpages = {14},
  year = {2015},
  month = {Aug},
  publisher = {American Physical Society},
  doiOPT = {10.1103/PhysRevB.92.064510},
  url = {https://link.aps.org/doi/10.1103/PhysRevB.92.064510}
}

@article{Happacher:2013pd,
  title = {Phase diagram of ${}^{4}$He on graphene},
  author = {Happacher, Jodok and Corboz, Philippe and Boninsegni, Massimo and Pollet, Lode},
  journal = {Physical Review B},
  volume = {87},
  issue = {9},
  pages = {094514},
  numpages = {5},
  year = {2013},
  month = {Mar},
  publisher = {American Physical Society},
  doi = {10.1103/PhysRevB.87.094514},
  url = {https://link.aps.org/doi/10.1103/PhysRevB.87.094514}
}

@misc{Buckingham:1967os,
 author = {Buckingham, A.~D.},
 doi = {10.1002/9780470143582.ch2},
 isbn = {9780470143582},
 journal = {Advanced Chemical Physics},
 pages = {107},
 publisher = {Wiley},
 title = {{P}ermanent and {I}nduced {M}olecular {M}oments and {L}ong‐{R}ange {I}ntermolecular {F}orces},
 url = {https://onlinelibrary.wiley.com/doi/10.1002/9780470143582.ch2},
 year = {1967}
}

@article{Deringer:2021ky,
 author = {Deringer, Volker L. and Bartók, Albert P. and Bernstein, Noam and Wilkins, David M. and Ceriotti, Michele and Csányi, Gábor},
 doi = {10.1021/acs.chemrev.1c00022},
 journal = {Chem. Rev.},
 number = {16},
 pages = {10073},
 publisher = {American Chemical Society (ACS)},
 title = {{G}aussian {P}rocess {R}egression for {M}aterials and {M}olecules},
 url = {https://pubs.acs.org/doi/10.1021/acs.chemrev.1c00022},
 volume = {121},
 year = {2021}
}

@inbook{Townsend2019,
  title = {Post-Hartree-Fock methods: configuration interaction,  many-body perturbation theory,  coupled-cluster theory},
  ISBN = {9780128136515},
  url = {http://dx.doi.org/10.1016/B978-0-12-813651-5.00003-6},
  DOI = {10.1016/b978-0-12-813651-5.00003-6},
  booktitle = {Mathematical Physics in Theoretical Chemistry},
  publisher = {Elsevier},
  author = {Townsend,  Jacob and Kirkland,  Justin K. and Vogiatzis,  Konstantinos D.},
  year = {2019},
  pages = {63–117}
}

@article{Potentialrepo,
	journal = {{Github Repository}},
	title = {\href{https://github.com/paulsphys/HeBz}{Helium-Benzene potentials}},
	author = {Sutirtha {Paul}},
	url = {https://doi.org/10.5281/zenodo.17982887},
	year = {2024},
	doi = {10.5281/zenodo.17982887}
}

@article{paperrepo,
	journal = {\url{https://github.com/DelMaestroGroup/papers-code-HeBenzene}},
	title = {Github repository},
	author = {Shahzad {Akram}, Sutirtha {Paul} and Adrian {Del Maestro}},
	url = {https://github.com/DelMaestroGroup/papers-code-HeBenzene},
	year = {2025},
	doi = {https://doi.org/10.5281/zenodo.17992891}
}

@article{datarepo,
	journal = {{Zenodo Data Repository}},
	title = {{Raw QMC Data for Accurate Helium-Benzene Potential: from CCSD(T) to Gaussian Process Regression}},
	author = {Sutirtha Paul and Adrian {Del Maestro}},
	url = {https://zenodo.org/records/18049741},
	year = {2025},
	doi = {10.5281/zenodo.18049741}}
\end{spacing}
\end{document}